\documentclass[fleqn,usenatbib]{mnras}

\usepackage{newtxtext,newtxmath}

\usepackage[T1]{fontenc}

\DeclareRobustCommand{\VAN}[3]{#2}
\let\VANthebibliography\thebibliography
\def\thebibliography{\DeclareRobustCommand{\VAN}[3]{##3}\VANthebibliography}

\usepackage{subfigure}
\usepackage{graphicx,booktabs,multirow}
\usepackage[utf8]{inputenc}
\usepackage[T1]{fontenc}
\usepackage{booktabs}
\usepackage{float} 
\usepackage{lineno} 
\usepackage{placeins}
\usepackage{longtable}
\usepackage{orcidlink}
\usepackage{multirow,array}
\usepackage{amsmath}
\usepackage{threeparttable}             
\usepackage{adjustbox}    
\usepackage{float}

\title{A comprehensive study of time delay between optical/near-infrared and X-ray emissions in black hole X-ray binaries}
\author[D. Du et al.]{
Dizhan Du$^{1}$,
Bei You\,\orcidlink{0000-0002-8231-063X}$^{2}$\thanks{E-mail: youbei@whu.edu.cn (BY)},
Zhen Yan\,\orcidlink{0000-0002-5385-9586}$^{3}$,
Xinwu Cao$^{1}$\thanks{E-mail: xwcao@zju.edu.cn (XWC)},
Jean-Marie Hameury$^{4}$,
Yue Wu\,\orcidlink{0000-0002-5225-637X}$^{5}$
\\
$^{1}$Institute for Astronomy, School of Physics, Zhejiang University, Hangzhou 310058, China\\
$^{2}$Department of Astronomy, School of Physics and Technology, Wuhan University, Wuhan 430072, China\\
$^{3}$Shanghai Astronomical Observatory, Chinese Academy of Sciences (CAS), Shanghai 200030, China\\
$^{4}$Universit\'{e} de Strasbourg, CNRS, Observatoire Astronomique de Strasbourg, 67000 Strasbourg, France\\
$^{5}$School of Astronomy and Space Science, Nanjing University, Nanjing 210023, China
}

\date{Accepted XXX. Received YYY; in original form ZZZ}

\pubyear{\the\year{}}

\newcolumntype{C}[1]{>{\centering\arraybackslash}p{#1}}
\newcolumntype{R}[1]{>{\raggedleft\arraybackslash$}p{#1}<{$}}
\begin{document}
\label{firstpage}
\pagerange{\pageref{firstpage}--\pageref{lastpage}}
\maketitle

\begin{abstract}
We conducted a comprehensive study of daily delays using multi-wavelength data from a sample of well-studied black hole X-ray binaries, specifically focusing on the sources GX 339-4, 4U 1543-47, and XTE J1550-564. The Interpolated Cross-Correlation Function method was employed to investigate the temporal relationship between the X-ray (Compton component) and optical-infrared (OIR) emissions. Our results show that during the rising hard state, the Compton emission consistently lags behind OIR emission for several days. In contrast, during the decaying hard state, the OIR emission lags behind the Compton emission by approximately 6 to 35 days. This measurement can potentially be used in models of accretion physics and disk instability. We explore the underlying mechanisms responsible for these time delays, highlighting the critical role of viscous heating in the accretion disk in generating OIR luminosity for these sources. The observed time delays during both the rising and decaying hard states are well explained by the disk instability model.
\end{abstract}




\section{Introduction}

A black hole X-ray binary (BHXRB) is a binary system that consists of a black hole (BH) and a normal star, in which matter from the companion star is accreted and forms a disk around the black hole. Transient BHXRBs undergo dramatic X-ray, optical, and radio outbursts, separated by long periods of quiescence that can last from years to decades, or even longer \citep{mcclintock_black_2006}. There also exist transient X-ray binaries containing a neutron star, but their prevalence as compared to black hole systems is significantly less than for persistent systems
\citep{king_black_1996}.

The disk instability model (DIM) was developed to explain the outbursts observed in compact binaries \citep[see][and references therein]{hameury_review_2020}. During the quiescent phase of a binary system, the accretion disk slowly builds up; its temperature increases until it reaches at some place in the disk the hydrogen ionization temperature, which triggers a thermal-viscous instability and results in an outburst.
In the outburst phase, the light curves of BHXRB systems display a variety of profiles, the most common being the fast-rise exponential-decay type observed in X-rays and optical-infrared (OIR)  \citep{harmon_batse_1994,tanaka_black_1995}. Other patterns can be observed, including plateaus, multiple peaks, secondary maxima \citep{chen_properties_1997,Yan2015, tetarenko_watchdog_2016}.
Typically, an X-ray outburst is accompanied by multi-wavelength flares, ranging from radio to gamma-rays \citep{mcclintock_black_2006,you_observations_2023}. It is important to note that different emitting regions and processes are involved at different wavelengths.

Radio emission mainly comes from the synchrotron process occurring in jets \citep{fender_jets_2006, bright2020}, whereas hard X-ray emission generated by the scattering of soft X-rays within the Comptonizing corona.\citep{narayan_advection-dominated_1998,poutanen_accretion_1998,Kawamura2023,you_observations_2023,you2024} and/or relativistic jets \citep{markoff_jet_2001,kara2019,ma2021,marino2021,you2021,peng2023,zhang2023}. During the decaying hard state of MAXI J1820+070 (while transitioning from the soft state back toward the quiescent state), it was observed for the first time that the radio flux lags behind the X-ray flux by approximately 8 days \citep{you_observations_2023}. These day-scale delays are hardly explained by jet propagation but can be naturally accounted for by an expanding advection-dominated accretion flow (ADAF). In this scenario, the magnetic field strength within the expanding ADAF lags behind the hard X-ray emission. Since jet power is linked to the inner magnetic field strength, this delay naturally produces the observed radio lag. This implies that the X-ray emission during the decaying hard state of MAXI J1820+070 is primarily produced in the ADAF.

As for the optical bands, three possible main mechanisms have been proposed to explain the observed emission: X-ray reprocessing, viscous heating, and jets. Additionally, hot accretion flows may also contribute to the optical emission. Their relative contributions are unclear, vary with time (and spectral state), and probably also depend on sources. Moreover, whereas the jet contribution likely dominates the mid-infrared, its contribution can be small in the near infrared and the visible domains \citep[see, e.g.][]{kosenkov_colors_2020}.
It has been proposed that optical emission primarily arises from the outer accretion disk due to X-ray reprocessing \citep{cunningham_returning_1976,vrtilek_observations_1990,van_paradijs_absolute_1994,van_paradijs_accretion_1996,russell_global_2006}. X-ray irradiation from the inner accretion flow, e.g., ADAF, can illuminate and heat the outer accretion disk, which prevents hydrogen from recombining. Consequently, the outer disk can be kept in a hot state, and the outburst lasts longer. 
Whereas irradiation dominates over viscous heating at radii larger than $10^4 R_G$ in steady state \citep{bollimpalli_disc_2018}, this need not be true at all times in time-dependent disks. Thermal emissions from the outer accretion disk, heated by viscosity, could also contribute significantly to the OIR emission \citep{shakura_black_1973,frank_accretion_2002}.
In addition, it was also argued that the flat optically thick spectrum of the jets extends from the radio to the OIR 
\citep{corbel_near-infrared_2002,chaty_multiwavelength_2003,markoff_exploring_2003,brocksopp_soft_2004,homan_multiwavelength_2005,fender_jets_2006,russell_global_2006,buxton_optical_2012,tetarenko_sub-mm_2015,fender_comprehensive_2023,john_correlated_2024,mastroserio_x-ray_2025}.

The correlations between the radio, optical, and X-ray emissions have often been used to identify the dominant emission processes in the OIR spectrum. \cite{russell_global_2006} presented a global correlation between OIR and X-ray luminosities for BHXRBs in the hard X-ray state, expressed as \(L_{\rm OIR} \propto L_{\rm X}^{0.6}\). This correlation can be interpreted as resulting from either jet emission or X-ray reprocessing in the accretion disk. 
For GX 339-4, tight correlations between OIR and X-ray emissions were observed across four decades in X-ray luminosity, both in hard states, with a notable break in the correlation in the hard state \citep{coriat_infraredx-ray_2009}; one can also note the $L_{\rm X} - L_{\rm opt}$ relation is not the same in the hard and soft state. The differences observed between the optical/X-ray and IR/X-ray correlations suggest that the jet primarily contributes to the near-infrared emission during the hard state, while the optical emission is likely dominated by blackbody radiation from the accretion disk in both hard and soft states.

Timing analysis of the OIR variability in the short timescale of seconds was also employed to study the origin of OIR emission. Fast OIR variability in GX 339-4 was observed in August 2008, June 2007, and September 2015 \citep{casella_fast_2010, gandhi_rapid_2010, vincentelli_physical_2019}. This rapid variability cannot be explained by standard disk reprocessing, as the OIR lags behind the X-ray emission by approximately 0.1 second. Instead, it was proposed by e.g. \cite{vincentelli_physical_2019}, that the fast OIR variability originates from the jet, indicating that a portion of the OIR emission likely comes from jet activity.

Regarding the OIR variability on timescales of days, the inter-band time delay is also crucial for understanding the origin of the OIR, but is rarely explored. \cite{you_observations_2023} focused on multi-wavelength emissions, including radio, OIR, and X-ray, during the 2018 main outburst of MAXI J1820-070. The Interpolated Cross-Correlation Function (ICCF) analysis between the X-ray and OIR light curves during the decaying hard state revealed an unprecedented optical delay of approximately 17 days relative to the X-ray emission. This delay shows that the observed optical emission is not the result of X-ray reprocessing in the disk. They also found that the optical emission lags behind the radio emission from the jet by about 9 days, further indicating that the optical emission is unlikely to originate from the jet. 
Instead, these observed optical delays suggest that the optical emission primarily originates from the viscously heated disk, with its time-dependent variation being explained by the DIM. This conclusion was further supported by directly modeling the observed optical light curve using numerical simulations of the DIM \citep{you_observations_2023}. 
Note that the spectral energy distribution (SED) fitting analysis suggests an additional power-law component in the UV/optical/NIR frequency ranges during rebrightening events following the 2018 main outburst \citep{ozbey_arabaci_multiwavelength_2022,yoshitake2022}. This may indicate a partial contribution from a jet component to the overall OIR emission.

MAXI J1820+070 is not the only source being observed across multiple wavelengths with high cadence. During the era of the Rossi X-ray Timing Explorer (RXTE), outbursts of numerous BHXRBs were documented \citep{dunn_global_2010,dunn_global_2011}, some of which were detected in OIR bands \citep{jain_multiwavelength_2001,buxton_2002_2004}. However, the cross-correlation between the X-ray and OIR emissions has not been thoroughly investigated. Here, we conduct a comprehensive study of time delays between X-ray and OIR emissions for a sample of RXTE BHXRBs, aiming to understand the origin of the OIR emission. This analysis also provides insights into the DIM and the role of disk winds. 

We outline our sample selection of BHXRBs and data reduction in Section \ref{sec:sample}. The model selection, spectrum fitting, and X-ray luminosity calculations are detailed in Section \ref{sec:fitting}. Section \ref{sec:ccf} presents the ICCF method used to quantitatively analyze the time delay between OIR and X-ray emissions, along with the results. Finally, we present our discussions in Section \ref{sec:conclusion} and present a summary in Section \ref{sec:sum}. For clarity and ease of reading, only figures corresponding to the GX 339-4 2002-2003 outburst are included in the main text; those for other outbursts of GX 339-4 and other sources are provided in the Appendix.

\section{Sample selection and Data reduction}\label{sec:sample}

RXTE accumulated extensive raw observational data through its 17-year monitoring campaign of BHXRBs. Given that the aim of this work is to explore the time delay between OIR and X-ray, only sources with high-cadence X-ray and OIR coverage in the hard state are considered. We found that three sources (XTE J1550$-$564, GX 339-4, and 4U 1543$-$47) satisfy this requirement with available data in the literature. Table \ref{tab:properties} lists the relevant information for these three BHXRBs.

GX 339$-$4 is a Galactic low-mass X-ray binary that was first observed during an X-ray outburst in 1972 \citep{markert_observations_1973}. The source underwent four outbursts from 2002 to 2011, details on the outbursts and state transitions can be found in \cite{belloni_evolution_2005}, \cite{belloni_integralrxte_2006}, \cite{del_santo_broad-band_2009}, \cite{marcel_unified_2019}. 

XTE J1550$-$564 was first detected by the All-Sky Monitor (ASM) aboard RXTE in September 1998 \citep{smith_xte_1998}, details on the 2000 outburst and state transitions can be found in \cite{corbel_x-ray_2001}. 

The recurrent X-ray transient 4U 1543$-$475 was first discovered in 1971 \citep{matilsky_new_1972}. Details on the 2002 outburst and state transitions can be found in \cite{kalemci_multiwavelength_2005}.

We used all RXTE/PCA archival data for each source. The good time intervals for each observation are produced by running the task \texttt{maketime}. The source and background spectra are produced by the script \texttt{pcaextspec2}, and the spectra are grouped with a minimum of 25 counts per bin. 

The OIR data for XTE J1550-564 and 4U 1543-564 were obtained through observations using the YALO telescope \citep{jain_multiwavelength_2001,buxton_2002_2004}. We used the \texttt{Origin}\footnote{\url{https://www.originlab.com/}} software to extract data from light curves where the data themselves were unattainable. The OIR data for GX 339-4 was taken from SMARTS \footnote{\url{http://www.astro.yale.edu/smarts/xrb/home.php\#}}. The photometric reduction procedure and results have been reported in \citet{buxton2012}. The VEGA system was used to convert magnitudes to flux, using zero-points of 3636, 1580, 1021, and 640 Jy and effective wavelengths of 545, 1220, 1630, and 2190 nm for the V, J, H, and K filters, respectively \citep{bessell_model_1998}.

The OIR flare emerges several weeks to months after the onset of the outburst decline.
To eliminate the flux contribution of the underlying exponential decay from the main outburst, we fitted the OIR luminosity light curves in a decaying hard state using an exponential plus Gaussian function, following the methodology outlined in \cite{you_observations_2023}. These adjusted light curves, with the exponential decay component subtracted, were subsequently used to estimate the lag between the X-ray and OIR flares.
\renewcommand{\arraystretch}{1.2}
\begin{table*}
\caption{Properties and data collected for the BHXBs.} 
\label{tab:properties}
\centering
\begin{tabular}{cccccccc}
\hline
\textbf{Source} & 
\parbox{1.5cm}{\centering Dist./kpc} & 
\parbox{1.5cm}{\centering Period/h} & 
\parbox{1.5cm}{\centering BH mass \(M_{1}  / \mathrm{M}_{\sun}\)} & 
\parbox{1.5cm}{\centering Mass ratio \(q=\frac{M_{2}}{M_{1}}\)} &
\multirow{2}{*}{\parbox{1.8cm}{\centering X-ray flare in decaying hard state}} & 
\multirow{2}{*}{\parbox{1.8cm}{\centering OIR flare in decaying hard state}} & 
\multirow{2}{*}{\parbox{1.8cm}{\centering OIR band data references}} \\
\textbf{} & 
\textbf{(ref)} & 
\textbf{(ref)} & 
\textbf{(ref)} & 
\textbf{(ref)} & 
& & \\
\hline
GX 339-4 & \(>5\) & \(42.21\) & 2.3--9.5 & \(\leq0.18\pm0.05\) & \multirow{2}{*}{\checkmark} & \multirow{2}{*}{\checkmark} & \multirow{2}{*}{SMART} \\         
= V821 Ara & (1,2) & (2,3,4) & (2) & (2) & & &  \\ 
\hline
XTE J1550-564 & \(4.5\pm0.5\) & \(37.01\) & 7.8--15.6 & \(\approx 0.03\) & \multirow{2}{*}{\checkmark} & \multirow{2}{*}{\checkmark} & \multirow{2}{*}{(10,11)} \\        
= V381 Nor & (5,6) & (5,6) & (5,6) & (5,6) & & & \\
\hline
4U 1543-47 & \(7.5\pm0.5\) & \(26.79\) & 8.4--10.4 & 0.25--0.31 & \multirow{2}{*}{\checkmark} & \multirow{2}{*}{\checkmark} & \multirow{2}{*}{(12)} \\         
= IL Lup & (7) & (8,9) & (8,9) & (8,9) & & & \\ 
\hline
\end{tabular}

\vspace{1ex}
\noindent\textbf{References:} 
(1)~\cite{hynes_distance_2004}, 
(2)~\cite{heida_mass_2017}, 
(3)~\cite{hynes_dynamical_2003}, 
(4)~\cite{levine_detection_2006}, 
(5)~\cite{orosz_improved_2011}, 
(6)~\cite{orosz_dynamical_2002}, 
(7)~\cite{jonker_distances_2004}, 
(8)~\cite{orosz_orbital_1998}, 
(9)~\cite{orosz_inventory_2003}, 
(10)~\cite{jain_multiwavelength_2001}, 
(11)~\cite{jain_optical_2001}, 
(12)~\cite{buxton_2002_2004}
\end{table*}
\renewcommand{\arraystretch}{1.0}

\begin{figure*}
    
    \centering  
    \includegraphics[width=0.9\textwidth]{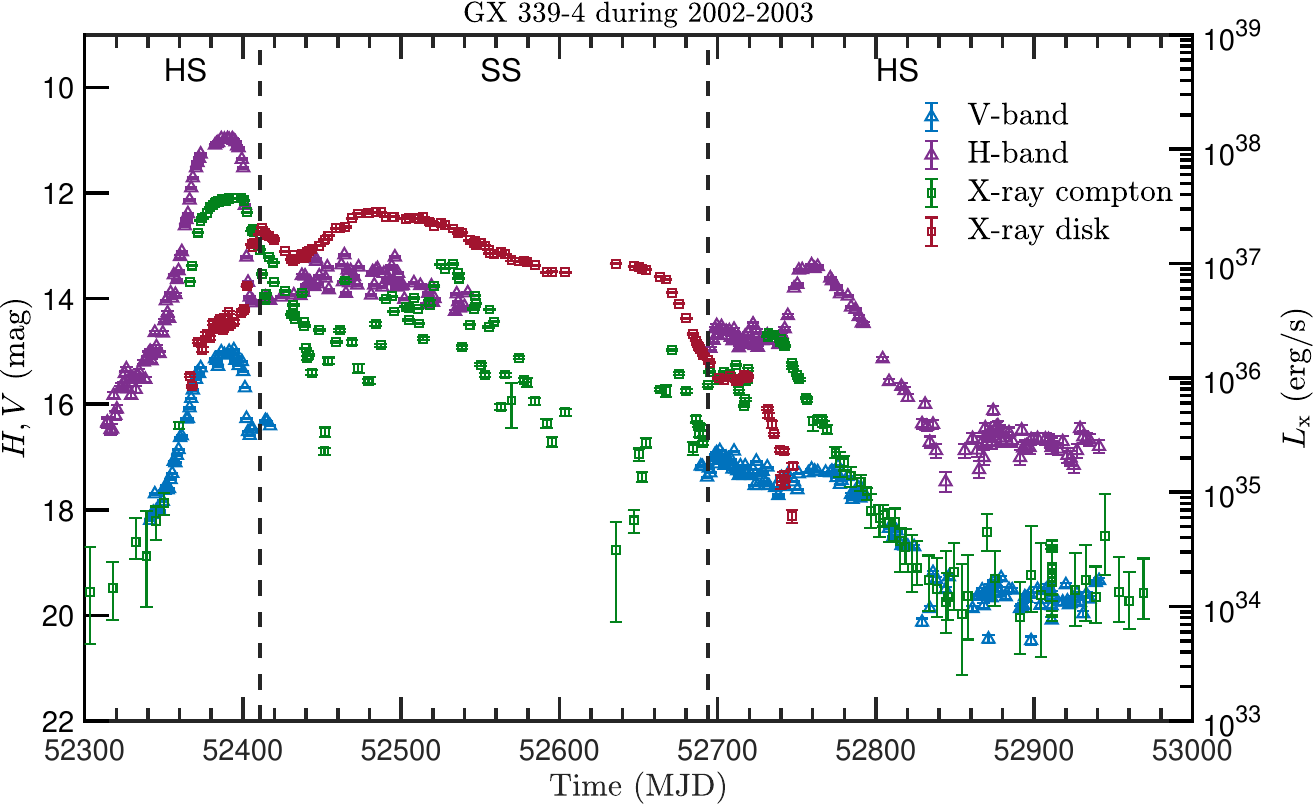}  
    \caption{Multi-wavelengths lightcurves of GX339-4 from MJD 52300 to 53000. Dashed lines correspond to transitions between Hard (HS) and Soft (SS) states.}  
    \label{fig:339_1}  
\end{figure*} 

\section{X-ray Spectra fitting} \label{sec:fitting}
To analyze the energy spectra in the 3–25 keV range, we employed the {\tt XSPEC} package (version 12.10.1), applying a 1\% systematic error to the data. The spectral fitting was performed for individual OBSIDs, with exposure times ranging from 16 to 17888 s.

The spectral fitting was performed using the model {\tt TBabs*(diskbb + gaussian + nthcomp)} in GX 339-4. In this model, the {\tt diskbb} component represents the multi-temperature blackbody emission from the accretion disk, while the {\tt nthcomp} component accounts for the inverse Comptonization processes occurring in the hot corona \citep{zdziarski_broad-band_1996, zycki_1989_1999}. Due to the limited energy range of the RXTE spectra (3--25 keV), the electron temperature in the corona could not be precisely constrained and was therefore fixed at 60 keV \citep{you_x-ray_2023}. The hydrogen column density was set to \(N_{\rm H} = 4 \times 10^{21}~\rm cm^{-2}\) \citep{done_re-analysis_2010}.

The {\tt gaussian} component was included to model the iron emission line, centered at 6.4 keV, with the line width treated as a free parameter. For the 2006–2007 outburst (MJD 54000–54450), the energy spectra were well-fitted using only the Comptonization component during the hard state, rendering the disk component unnecessary. The best-fit results were consistent with those reported in previous studies \citep{dunn_global_2010}, where the hard component was fitted by a powerlaw instead of {\tt nthcomp}.

In the case of XTE J1550-564, we employed the model \( \texttt{TBabs}*\texttt{smedge(diskbb + nthcomp + gaussian)} \). The inclusion of a smeared iron absorption edge near 8 keV significantly improved the fits, along with the presence of a Fe emission line centered at 6.5 keV with a fixed width of 1.2 keV \citep{sobczak_x-ray_1999,ebisawa_spectral_1994}. The hydrogen column density was fixed at \( N_{\rm H} = 2 \times 10^{22}~\rm cm^{-2} \) \citep{sobczak_x-ray_1999}. The same \( \texttt{nthcomp} \) and \( \texttt{gaussian} \) components were applied as in the analysis of GX 339-4. The best-fitting results are consistent with those reported in previous studies\citep{sobczak_x-ray_1999}, where the hard component was fitted by a powerlaw instead of {\tt nthcomp}.

The spectral and temporal evolution of 4U 1543-475 was analyzed using different models depending on the source state. We used the models \( \texttt{TBabs*(diskbb+nthcomp+gaussian)} \) in the soft state and \( \texttt{TBabs*nthcomp} \) in the hard state. The hydrogen column density was fixed at \( N_{\rm H} = 3.8\times 10^{21}~\rm cm^{-2} \) \citep{kalemci_multiwavelength_2005}. The best-fitting results are consistent with those reported in earlier studies \citep{park_spectral_2004}, where the hard component was fitted by a powerlaw instead of {\tt nthcomp}.

The \( \texttt{cflux} \) command in XSPEC was used to estimate the unabsorbed disk thermal luminosity and Comptonization luminosity in the 3-25 keV energy range. Data points with luminosity errors exceeding \(5\sigma\), primarily due to low flux levels and insufficient fitting constraints, were excluded. The light curves in X-rays and OIR bands for GX 339-4, XTE J1550-564, and 4U 1543-47 are presented in Fig.\ref{fig:339_1}, Figs.\ref{fig:339_2}-\ref{fig:339_4}, Fig.\ref{fig:1550}, and Fig.\ref{fig:1543}, respectively. The spectral fit parameters for these three sources are shown in Tables~\ref{tab:339spectral}-\ref{tab:1543spectral}; an example of spectral fitting is shown in Fig.\ref{fig:339fitting}.

\section{Cross-Correlations between optical/NIR and X-ray} \label{sec:ccf}
\renewcommand{\arraystretch}{1.35} 
\begin{table*}
\caption{Time lags between different optical wavelengths and Compton luminosity using two methods}
\label{tab:lag}
\centering
\begin{tabular}{C{2.0cm} C{2.0cm} C{1.5cm} R{2.3cm} R{2.3cm} R{2.3cm} R{2.3cm}}
\hline
Source & MJD & OIR band & \multicolumn{2}{c}{Rise} & \multicolumn{2}{c}{Decay} \\
       & (d) &          & \text{Centroid (d)} & \text{Peak (d)} & \text{Centroid (d)} & \text{Peak (d)} \\
\hline
\multirow{8}{*}{GX 339-4} 
& \multirow{2}{*}{52300--53000} & H & -2.41^{+0.99}_{-0.85} & -3.10^{+0.50}_{-0.70} & 25.32^{+1.65}_{-1.06} & 25.26^{+1.80}_{-1.56} \\
& & V & -2.09^{+0.95}_{-0.67} & -2.10^{+0.40}_{-0.80} & 34.64^{+1.88}_{-1.77} & 35.35^{+3.36}_{-3.22} \\
\cline{2-7}
& \multirow{2}{*}{53000--53700} & H & -1.72^{+1.71}_{-1.09} & -3.24^{+1.72}_{-0.92} & 20.20^{+3.46}_{-1.75} & 20.13^{+1.98}_{-1.59} \\
& & V & -5.68^{+0.74}_{-0.81} & -5.72^{+0.92}_{-1.08} & 23.99^{+5.86}_{-1.70} & 23.64^{+6.54}_{-2.16} \\
\cline{2-7}
& \multirow{2}{*}{54050--54450} & H & -2.54^{+0.91}_{-0.85} & -3.20^{+0.55}_{-0.50} & {-} & {-} \\
& & V & -1.87^{+0.88}_{-0.68} & -3.00^{+0.55}_{-0.70} & {-} & {-} \\
\cline{2-7}
& \multirow{2}{*}{55200--55800} & H & -5.61^{+0.50}_{-0.55} & -5.95^{+0.35}_{-0.70} & 17.86^{+1.44}_{-2.69} & 17.46^{+2.16}_{-2.10} \\
& & V & -6.38^{+0.45}_{-0.47} & -7.60^{+0.60}_{-0.35} & 18.50^{+1.60}_{-2.92} & 15.96^{+4.34}_{-0.84} \\
\hline
\multirow{2}{*}{4U 1543-47} 
& \multirow{2}{*}{52400--52600} & J & {-} & {-} & 7.02^{+0.40}_{-0.42} & 6.78^{+0.90}_{-1.44} \\
& & K & {-} & {-} & 6.95^{+0.50}_{-0.43} & 6.60^{+1.40}_{-1.50} \\
\hline
\multirow{2}{*}{XTE J1550-564} 
& \multirow{2}{*}{51600--51800} & H & -6.67^{+2.11}_{-0.54} & -6.75^{+1.85}_{-0.30} & 25.74^{+1.10}_{-0.60} & 23.03^{+0.72}_{-0.78} \\
& & V & -3.53^{+1.29}_{-1.94} & -3.75^{+1.85}_{-1.10} & 30.65^{+2.18}_{-2.13} & 26.25^{+1.33}_{-1.05} \\
\hline
\end{tabular}
\end{table*}
\renewcommand{\arraystretch}{1.0}

ICCF is a powerful tool for analyzing time delays between different time-series signals. It is widely used in the time-series analysis of reverberation mapping (RM) in active galactic nuclei (AGNs) \citep{kaspi_reverberation_2000,bentz_lick_2009}. ICCF analysis has been applied to study time lags between different wavelengths in MAXI J1820+070 \citep{you_observations_2023}. For a detailed discussion of ICCF, the reader is referred to \cite{du_supermassive_2014} and \cite{you_observations_2023}. In this work, ICCF is employed to calculate the time delay between OIR and X-ray emissions during both the rising and decaying hard states of the three BHXRBs.

\subsection{Rising hard state analysis}
In the rising and decaying hard states, the Compton component exhibits significant flares, whereas the disk component shows only a monotonic rise or decay. Thus, we consider the OIR flare to be related to the Compton flare. In the \cite{you_observations_2023}, the ICCF analysis was performed between the Compton component and OIR as well.

In the rising hard state, the Compton component consistently lags behind the OIR by a few days, as shown in Table \ref{tab:lag}.
For GX 339-4, the X-ray delay ranges from 2 to 8 days, the 2002-2003 outburst lightcurves and details of the ICCF analysis are shown in Fig.\ref{fig:i339_1}(a), (c), and (e). The lightcurves of other GX 339-4 outbursts and details of the ICCF analysis are shown in Figs.\ref{fig:i339_2}(a)-\ref{fig:i339_4}(a), Figs.\ref{fig:i339_2}(c)-\ref{fig:i339_4}(c) and Figs.\ref{fig:i339_2}(e)-\ref{fig:i339_4}(e). During the 2004-2005 outburst of GX 339-4, two peaks in OIR and Compton luminosity were observed in the rising hard state (see Fig.~\ref{fig:i339_2}(a)), which is different from the behavior seen during other outbursts from this source. Since we care about the ascent to the peak from the quiescent state, the following analysis considers only the first peak. Nevertheless, it is worthwhile to note that the second Compton flare also lags the second OIR flare by a time comparable to that of the first pair of flares, if one calculates the lag from the ICCF of the full light curves. For the H band, the centroid and peak lags are -6.17 d and -4.40 d, respectively, while for the V band, the centroid and peak lags are -3.88 d and -4.85 d.

For XTE J1550-564, the X-ray delay ranges from 3 to 7 days. The light curves and ICCF analysis for this system are shown in Fig.\ref{fig:i1550}(a), Fig.\ref{fig:i1550}(c), and Fig.\ref{fig:i1550}(e). 
For 4U 1543-47, since the rising hard state has not been detected, an ICCF analysis cannot be performed.

\subsection{Decaying hard state analysis}
In the decaying hard state, the OIR emission consistently lags behind the X-ray emission by a few days up to nearly 30 days, as shown in Table \ref{tab:lag}.

\begin{figure*}
	\centering
	 \includegraphics[width=0.95\textwidth]{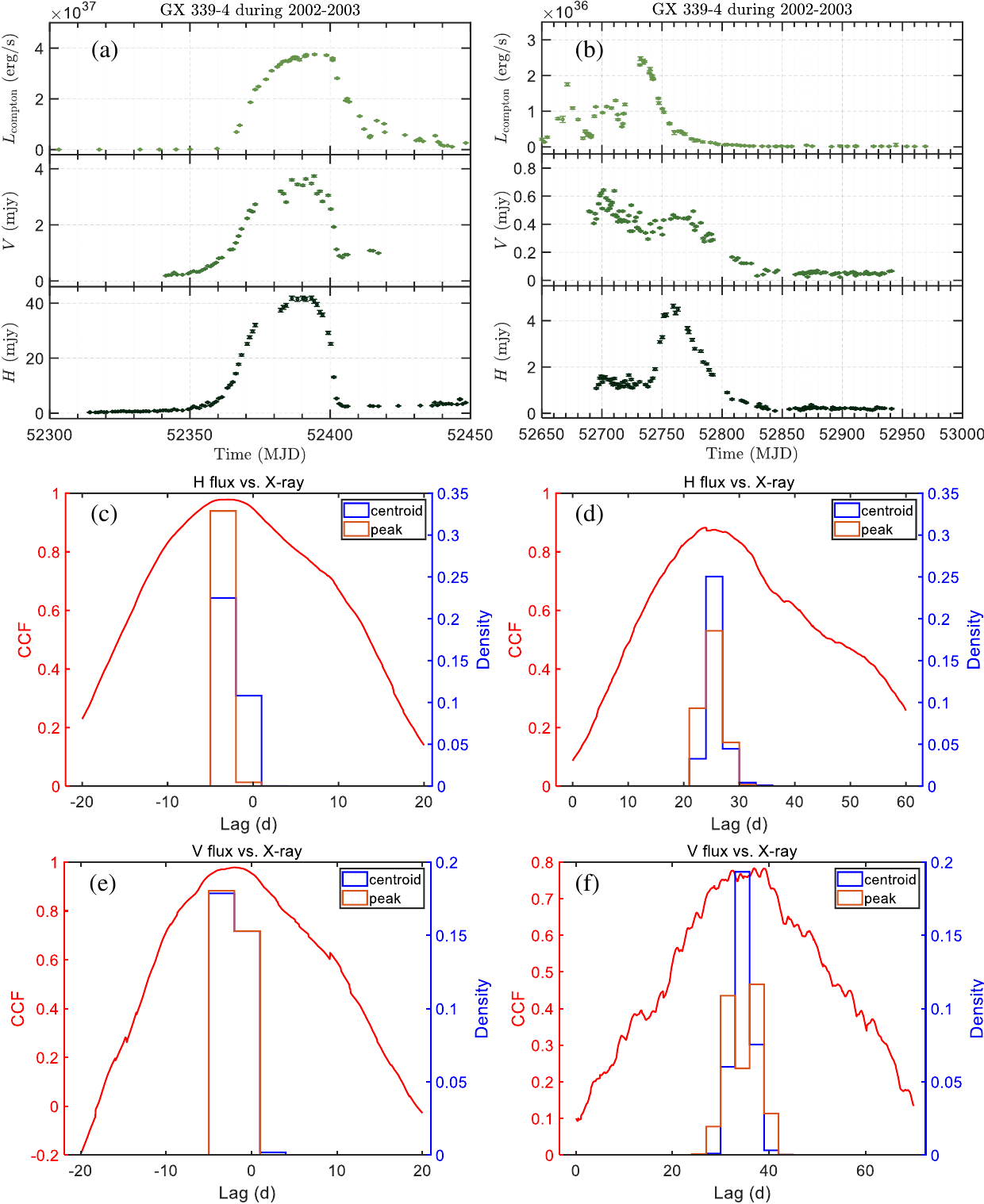}  
    	
\caption{
Panel a: Multi-wavelength monitoring during the rising hard state flare. Panel b: Multi-wavelength monitoring during the decaying hard state flare. Panel c: The cross-correlation analysis between the H-band and Compton X-ray luminosity, specifically before MJD = 52450 (red line). Panel d: Cross-correlation analysis between H-band and the Compton X-ray luminosity, after MJD = 52650 (red line). Panel e: The cross-correlation analysis between the V-band and Compton X-ray luminosity, specifically before MJD = 52450 (red line). Panel f: Cross-correlation analysis between V-band and the Compton X-ray luminosity, after MJD = 52650 (red line). In the four lower panels, the blue histograms show the distribution of cross-correlation centroid lags, determined from the centroid of ICCF above a threshold (\(r>0.8r_{\rm max}\), where $r_{\rm max}$ is the maximum correlation coefficient). The yellow histograms display the peak lags, corresponding to the time delays at which the maximum correlation occurs. The uncertainties in the lags are derived using the flux randomization/random subset sampling (FR/RSS) method. The corresponding axes are shown on the right. We performed 10,000 simulations using linear interpolation; details can be found in \citet{gaskell_accuracy_1987} and \citet{du_supermassive_2014}.
}
\label{fig:i339_1}
\end{figure*}

For GX 339-4, the 2002-2003 outburst exhibits a significant OIR delay, with a lag of nearly 25 days observed in the H band. Remarkably, the V band shows a delay of about 35 days, lagging the H band by approximately 10 days. The light curves and details of the ICCF analysis are presented in Fig.\ref{fig:i339_1}(b), (d), and (f). Note that the peak in V is much broader than in H, which can make it difficult to estimate the delay.

In the decaying hard state of the 2006–2007 outburst of GX 339-4, the Compton luminosity exhibited two flares roughly 80 days apart, yet only a single OIR flare was detected. Since the cause of the two flares in Compton luminosity remains unclear, the time delay may not reveal the true correlation. Therefore, we exclude this outburst from the ICCF analysis.

The lightcurves of other GX 339-4 outbursts and details of the ICCF analysis are shown in Figs.\ref{fig:i339_2}(b), (d), (f); and Figs.\ref{fig:i339_4}(b), (d), (f).

In the case of XTE J1550–564, no distinct reflare is observed in hard X-rays. The reflare begins around MJD 51670, coinciding with the intermediate or soft state \citep{homan_correlated_2001, corbel_x-ray_2001}. 

The OIR delay is approximately 23–30 days. The light curves and details of the ICCF analysis for this system are illustrated in Fig.\ref{fig:i1550}(b), Fig.\ref{fig:i1550}(d), and Fig.\ref{fig:i1550}(f).

For 4U 1543-47, the OIR delay is nearly 7 days. The lightcurves and details of the ICCF analysis for this system are shown in Fig.\ref{fig:i1543}(b), Fig.\ref{fig:i1543}(d), and Fig.\ref{fig:i1543}(f). Because of the scarcity of data points in the B, V, and I bands, and since rebrightening events were less pronounced in these bands, we performed the ICCF analysis exclusively for the J and K bands (see Table.\ref{tab:lag} and Fig.\ref{fig:i1543}). 

\section{Discussion} \label{sec:conclusion}

\subsection{OIR delay during the decaying hard state}

\subsubsection{Reflares in the context of the DIM}

As discussed in Section \ref{sec:ccf}, during the decaying hard state, the OIR emission lags behind the Compton component by up to 35 days. This behavior aligns with observations of MAXI J1820+070 reported in \cite{you_observations_2023}, suggesting that the OIR delay relative to Compton emissions is a common phenomenon in the decaying hard states of BHXRBs.

In MAXI J1820+070, the DIM incorporating disk winds successfully explains the observed delay of approximately 17 days between optical and X-rays \citep{you_observations_2023}. As the outburst transitions to the decaying hard state, assuming the truncated disk model, the expansion of the truncation radius, combined with the competing effects of a decreasing accretion rate, leads to the reflare of the Compton component \citep{you_observations_2023}. Meanwhile, the outer regions of the accretion disk are in the cold state with temperatures below the hydrogen ionization threshold. Part of this cold region is reheated by the Compton reflare, triggering a transition back to the hot state and producing the OIR rebrightening. As a result, hydrogen becomes ionized, which triggers the revival of disk instability. Moreover, the disk winds, which were observationally reported \cite{Sanchez2020}, help to remove angular momentum from the disk, producing an effect similar to an increase in viscosity \citep{begelman_compton_1983}. Consequently, this leads to a brighter optical peak that appears later compared to situations that do not incorporate disk winds.

\FloatBarrier

To further investigate this phenomenon, we applied the DIM to simulate the OIR flares of the three BHXRBs studied in this work. We utilized the same code as described in \citet{you_observations_2023}, which models the thermal–viscous instability related to hydrogen ionization. This model allows the disk to switch between hot and cold states when the surface density exceeds critical thresholds. It employs different viscosity parameters for the hot and cold states, includes disk truncation due to evaporation, and takes into account X-ray irradiation (input from the observed hard X-ray luminosity) that heats the outer disk. Additionally, the model incorporates mass loss through winds, which removes angular momentum and effectively increases viscosity. This process enables the heating front to propagate further outward, resulting in delayed OIR rebrightening during outbursts.

One should note that, in the DIM framework, we are unable to quantify the contribution of the jet to the OIR emission and explain a fast OIR variability which is most likely related to a jet (see below). This also means that our lightcurves are diluted by the contribution from the jet, so that the amplitudes we obtain are somewhat overestimated, especially in the mid-infrared. It nevertheless remains that the contribution from the disk to the light curve can be dominant, and, as we shall see, can naturally reproduce the observed delays and light curves. In the following simulation work, since, relative to the V band, the H band is more diluted by the jet contribution (see Sec.~\ref{sec:bbsed}), we performed simulations only for the V band, as it provides a more accurate representation.

\begin{figure}
    \centering  
    \includegraphics[width=0.5\textwidth]{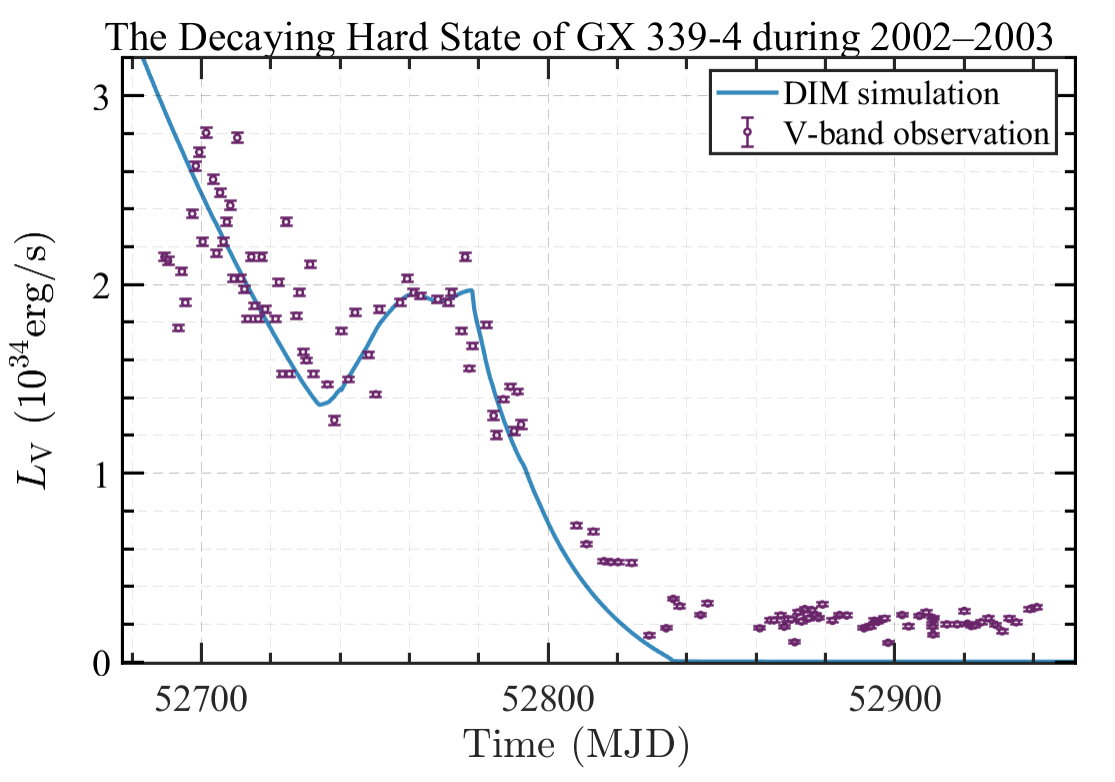}  
    \caption{Disk instability model with the disk wind simulation of GX 339-4 outburst from 2002 to 2003. The blue line represents the simulation results, and the purple dots indicate the observed data.}  
    \label{fig:339s1}  
\end{figure}

\begin{figure}
    \centering
    \includegraphics[width=0.9\columnwidth]{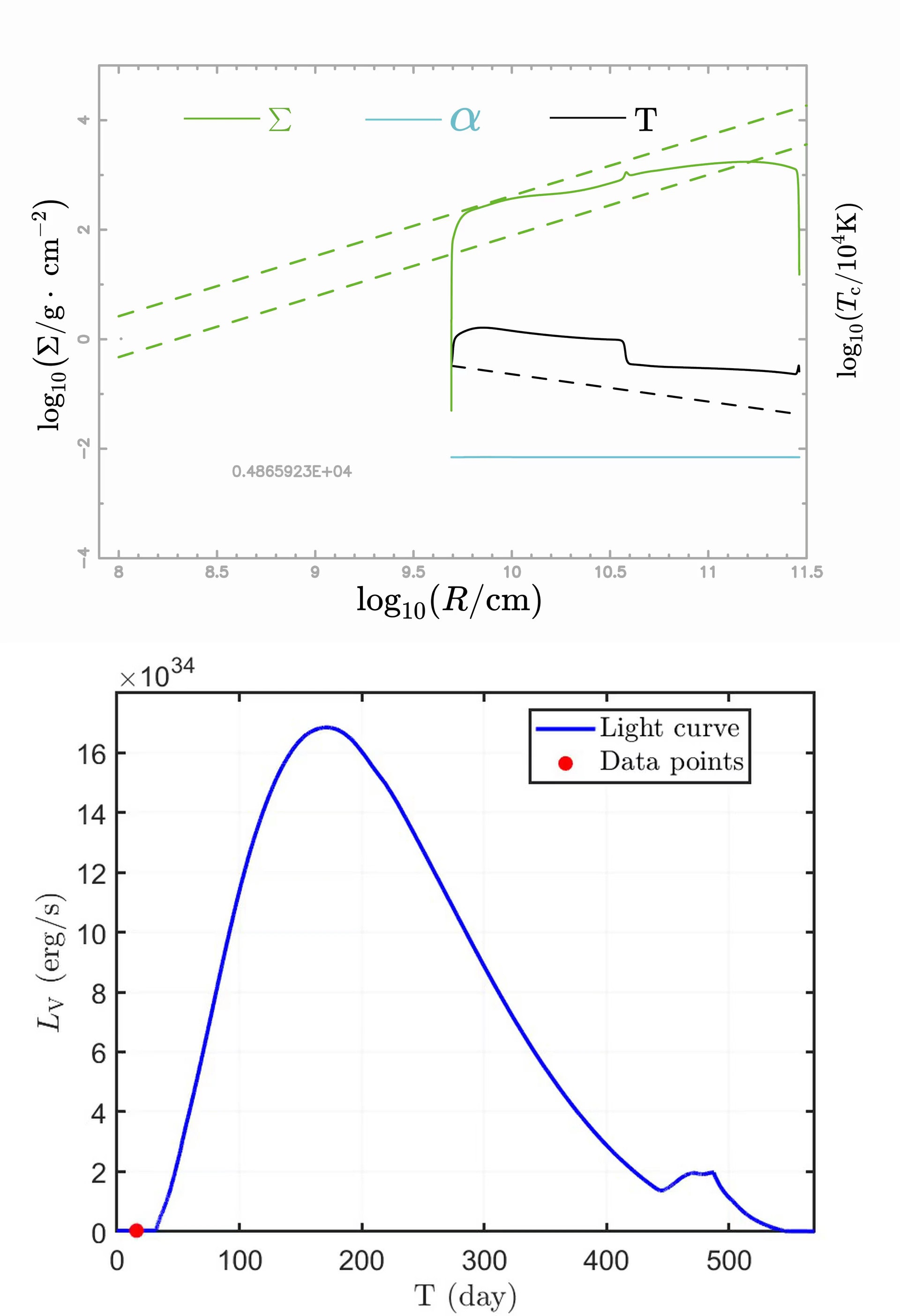}
    \caption{
    Representative frame from the animation showing the DIM evolution of the disk structure during the 2002–2003 outburst of GX 339-4.
    The top panel shows variations in surface density (green), central temperature (black), and the viscosity parameter \(\alpha\) (blue). The upper green dashed line denotes $\Sigma_{\rm max}$: in the cold state, if $\Sigma > \Sigma_{\rm max}$, the disk transitions to the hot state. Conversely, the lower green dashed line denotes $\Sigma_{\rm min}$: in the hot state, if $\Sigma < \Sigma_{\rm min}$, the disk transitions back to the cold state. The black dashed line represents heating of the disk temperature by Comptonized X-rays.
    The bottom panel marks the corresponding time position within the outburst (red dot).}
    An online animation is available as Supplementary material.
    
    \label{fig:animation_339_2002}
\end{figure}

For GX 339-4, the simulation results in the V-band are presented in Fig. \ref{fig:339s1}, Fig. \ref{fig:339s2}, and Fig. \ref{fig:339s4}. The corresponding evolution of the disk structure — showing variations in surface density and central temperature over time during the 2002–2003 outburst of 339-4 — is presented in Fig. \ref{fig:animation_339_2002}, which presents the beginning frame from the accompanying animation. The other two frames of the animation are shown in Fig.~\ref{fig:DIM_frame}. An online animation is available as Supplementary Material. The evolution of other outbursts are available as Supplementary material. Additional details and parameters can be found in Table \ref{tab:par}.
For the 2002 outburst of 4U 1543-47, the simulation results in the J-band are presented in Fig. \ref{fig:1543s}, additional details and parameters can be found in Table \ref{tab:par}.

The 2000 outburst of XTE J1550-564 lasted approximately 300 days. The OIR luminosity during this event was nearly two orders of magnitude lower than that of other BHXRBs, placing it outside the range of our simulations. One possible explanation for this faint outburst is that the instability was triggered when disk mass accumulation was minimal, preventing the heating front from reaching the outer edge of the disk, as predicted by several DIM simulations, particularly when the disk extends to large radii \citep[see e.g.][]{hl20}. 
Consequently, this outburst occurred shortly after the previous one in 1999 and was significantly fainter. 

The 2006–2007 outburst of GX 339-4, as discussed in Section \ref{sec:ccf}, exhibited two Compton flares, in contrast to a single, long-duration OIR flare. Notably, the second Compton flare showed no time delay relative to the OIR flare. We attribute the OIR flare to the DIM triggered by the first Compton flare. The decay of the OIR flare was interrupted by the onset of the second Compton flare, which extended its duration and enhanced its intensity. The simulation result in the V-band for this event is illustrated in Fig. \ref{fig:339s3}.

\begin{table*}
\centering
\caption{Simulation parameters}
\label{tab:par}
\resizebox{\textwidth}{!}{
\begin{tabular}{ccccccccccccccc}
\toprule
source & MJD & \(D\) & \(P\) & \(M_{\mathrm{BH}}\) & \(M_{\mathrm{star}}\) & \(\alpha _{h}\) & \(\alpha _{c}\) & \(\dot{M}_{\mathrm{t}}\) & \(\zeta _{\mathrm{d}}\) & \(\zeta _{\mathrm{c}}\) & \(f'_{\mathrm{max}}\) & \(g'\) & \(t_\mathrm{start}\) & \(t_\mathrm{end}\) \\
\midrule
\multirow{4}{*}{GX 339-4} & 52300-53000  & \multirow{4}{*}{7} & \multirow{4}{*}{42.21} & \multirow{4}{*}{6} & \multirow{4}{*}{1.08} & 0.0224 & 0.007 & 0.158 & $5\times 10^{-5}$ & $1.475\times 10^{-4}$ & 0.001 & $1.75\times 10^{-4}$ & 52728 & 52778 \\
 & 53000-53700 & & & & & 0.035 & 0.025 & 0.119 & $8\times 10^{-4}$ & $3.472\times 10^{-5}$ & 0.001 & $9.8\times 10^{-5}$ & 53464 & 53496 \\
 & 54050-54450 & & & & & 0.0224 & 0.007 & 0.158 & $5\times 10^{-5}$ & $1.135\times 10^{-4}$ & 0.001 & $2.62\times 10^{-4}$ & 54206 & 54316 \\
 & 55200-55800 & & & & & 0.0224 & 0.007 & 0.158 & $5\times 10^{-5}$ & $9.400\times 10^{-5}$ & 0.001 & $2.82\times 10^{-4}$ & 55602 & 55622 \\
\midrule
4U 1543-47 & 52480-52600 & 7 & 26.79 & 10 & 2.5 & 0.7000 & 0.050 & 20.00 & $1\times 10^{-5}$ & $4.500\times 10^{-4}$ & 0.01 & $1.00\times 10^{-3}$ & 52476 & 52487 \\
\bottomrule
\end{tabular}
}
\vspace{1mm}
\begin{minipage}{\textwidth}
\small
\( D \): distance (kpc); 
\( P \): orbital period (h); 
\( M_{\mathrm{BH}} \), \( M_{\mathrm{star}} \): black hole and companion star mass (in \(M_\odot\)); 
\(\alpha_{h,c}\): hot/cold state viscosity; 
\(\dot{M}_{\mathrm{t}}\): mass transfer rate from the donor (in \(10^{16}\,\mathrm{g\,s}^{-1}\)); 
\(\zeta_{\mathrm{d}},\zeta_{\mathrm{c}}\): irradiation efficiencies, based on observed Comptonized X-ray luminosity; 
\(f'_{\mathrm{max}}\): maximum local surface density loss factor (disk wind); 
\(g'\): specific angular momentum factor carried away by disk wind; 
\(t_{\mathrm{start}}, t_{\mathrm{end}}\): disk wind duration.
See \citet{you_observations_2023} for more details.
\end{minipage}
\end{table*}

To further interpret the time delay, we investigated the key physical factors that govern it. We found that the time delay is primarily affected by the luminosity and duration of Compton X-ray irradiation, as well as the mass distribution in the outer accretion disk and the disk size. During the 2002 outburst of GX 339-4, we observed a longer time lag compared to past outbursts. This is attributed to the fact that, while the luminosity of the Compton flare remained similar, the duration of the flare’s half-maximum luminosity in the decaying hard state was significantly longer in 2002. 
Our simulations suggest that prolonged irradiation allows the hot front to extend further outward. This, in turn, delays the emergence of the cooling front, leading to a later decay of the OIR flare and resulting in a longer time delay.

Furthermore, our simulation also revealed that the duration of OIR flares is influenced by the amount of matter present in the outer accretion disk, as well as the size of the disk itself. More mass in the outer disk allows the hot front to propagate further, which is limited by the disk’s size. This results in a longer duration for the OIR flare and an increased observed time delay. Additionally, the thermal disk wind plays a significant role in the OIR flare, with its generation and duration also affecting the time delay \citep{you_observations_2023}.


\subsubsection{Other scenarios for the origin of the OIR rebrightening}

Several alternative scenarios have been proposed to explain the delay of OIR rebrightening in relation to hard X-ray. \citet{dincer_x-ray_2012} and \citet{corbel_formation_2013} observed that the OIR rebrightening during the 2010–2011 outburst of GX 339-4 occurred approximately ten days after the onset of the power-law flux or X-ray count rate. 
\citet{dincer_x-ray_2012} suggested a scenario where irradiation of the secondary star takes place during the rise of the OIR rebrightening, while jet emissions dominate during the peak. Conversely, \citet{corbel_formation_2013} proposed that this delay represents the time required for the jets to evolve, starting with an optically thin spectrum at radio frequencies and gradually changing to an optically thick synchrotron emission in OIR band. This evolution is likely driven by increased density and particle acceleration along the extended jet region. 
The optically thin-to-thick jet model has also been applied to explain the rebrightening in OIR flux observed in MAXI J1836-194 \citep{russell_evolving_2013,russell_accretion-ejection_2014}, as well as the UV/OIR rebrightening of MAXI J1820+070 \citep{ozbey_arabaci_multiwavelength_2022,constanza2024} and XTE J1550-564 \citep{russell_evidence_2010,russell_tool_2011} during their decaying hard states. 

The rebrightening phenomenon was also investigated using color-magnitude diagrams (CMDs), such as the V versus V-H diagram \citep{russell_tool_2011, poutanen_colours_2014, kosenkov_colors_2020}. According to \cite{russell_tool_2011}, differentiate between disk and jet contributions during the 2000 outburst of XTE J1550–564, utilizing only two wavebands. The OIR data follow a blackbody-like curve, with deviations attributed to a non-thermal jet component that primarily dominates during the rising and decaying hard states but becomes less significant in the soft state. \cite{poutanen_colours_2014} attributed the OIR deviations of XTE J1550-564 to the hybrid hot flow model, wherein non-thermal electrons emit synchrotron radiation in the OIR band.

\subsection{On the X-ray delay during the rising hard state}

\begin{figure}
    \centering  
    \includegraphics[width=0.46\textwidth]{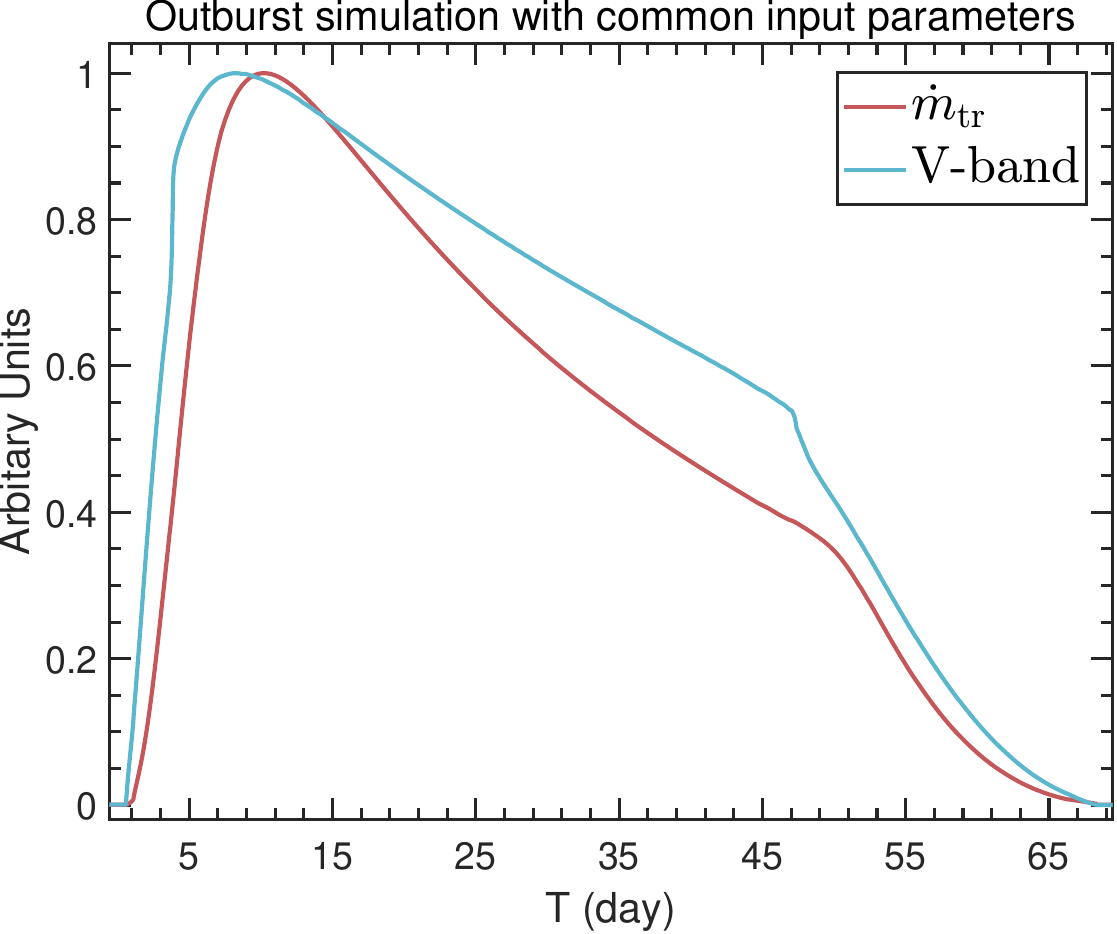}   
    \caption{An outburst simulation with common input parameters (\(\alpha _{h}=0.16,  \alpha _{c}=0.03,  \dot{M}_{\rm{t}}=10, \zeta _{\rm{d}}=8\times10^-4\)), the blue line represents \(\dot{m}_{\rm{tr}}\), the red line shows the visible flux in the V band. Both are normalized to unity at maximum.}
    \label{fig:rising sim}
\end{figure}

During the rising phase of an outburst (i.e., the rising hard state), we find that the OIR emission precedes the hard X-ray (Compton component) emission for the BHXRBs in this work. 
This phenomenon can be naturally explained by the DIM. The DIM predicts that the accretion rate at the truncation (inner disk) radius, \(\dot{M}_{\rm{tr}}\), lags behind the V-band light curve by a few days, as shown in Fig. \ref{fig:rising sim}. Meanwhile, since the Compton luminosity increases with increasing mass accretion rate at the truncation radius (e.g. it goes as \(L_{\rm{C}} \propto \dot{M}_{\rm{tr}}^2\) in the case of the ADAF model) \citep{narayan_advection-dominated_1995}, this leads to a delay in the Compton luminosity \(L_{\rm{C}}\) relative to the OIR emission.

At the onset of an outburst, a region close to the inner edge of the disk ionizes, the local temperature increases, and two heat fronts propagate inward and outward \citep{cannizzo_limit_1993,cannizzo_accretion_1993,hameury_review_2020}. Simultaneously, irradiation from the corona heats the disk, enhancing its luminosity \citep{cunningham_returning_1976,van_paradijs_absolute_1994}. 
As the heating front reaches the outer edge of the disk, the OIR emission soon attains its maximum. Due to viscous dissipation, matter flows through the entire disk towards the inner regions. As a consequence, the surface density decreases in the outer regions but continues to rise in the inner disk \citep[see the top right panel of Figure 5 in][]{dubus_disc_2001}. The accretion rate onto the black hole, therefore, keeps increasing before it reaches its peak. The delay in ${\dot{M}_{\rm tr}}$ relative to the OIR emission is thus significantly influenced by the viscosity in the hot state.

\begin{figure}
    \centering  
    \includegraphics[width=0.48\textwidth]{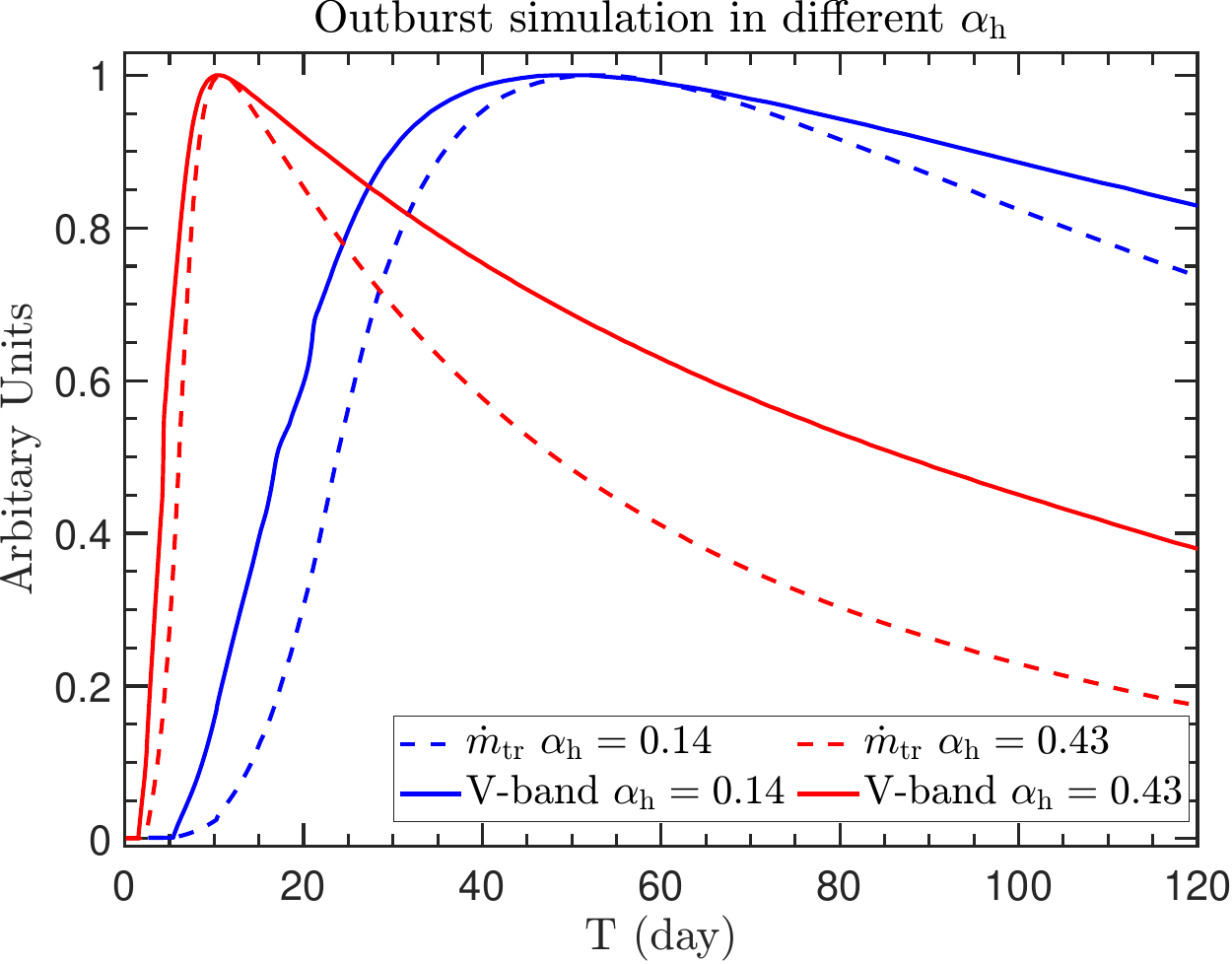}   
    \caption{An outburst simulation using the same common input parameters as in Fig. \ref{fig:rising sim}, but differing in \( \alpha_{\rm{h}} \), shows that a higher \( \alpha_{\rm{h}} \) results in a shorter time delay between the V-band and \( \dot{m}_{\rm{tr}} \).}
    \label{fig:rising2}
\end{figure}

To demonstrate the impact of viscosity on the resulting X-ray delay, we conduct simulations of the DIM for various values of hot-state viscosity, predicting variations in both the V-band flux from the outer disk and the Compton X-ray flux from the inner accretion flow within the truncation radius (see Fig. \ref{fig:rising2}). The results indicate that $\alpha_{\rm{h}}$ primarily influences the time delay between the V-band and $\dot{M}_{\rm{tr}}$, such that as $\alpha_{\rm{h}}$ increases, the lag of $\dot{M}_{\rm{tr}}$ behind the V-band decreases.

X-ray lagging behind UV to OIR has also been observed in other sources.
\cite{degenaar_multi-wavelength_2014} reported that in Swift J1910.2–0546, the X-ray flux decreased approximately six days after changes at UV to NIR wavelengths, before the source transitioned to the soft state. \cite{yan_ultraviolet_2012} found that during the 2010 outburst of GX 339-4, the hard X-ray dropped about ten days after the UV light started to fade. They attributed the lag to optically thick synchrotron emission from the compact jet. The 2010 outburst of GX 339-4 was accompanied by radio observations that tracked the rising hard state \citep{corbel_universal_2013}. ICCF analysis reveals that OIR emissions precede radio emissions by approximately 2 days. Whether the OIR emission originates from the jet remains open to discussion.

Besides the delays derived from the ICCF analysis in entire rising phase, the onset of X-ray outbursts, compared to the OIR lighcurve, was also considered in the literature to study the delay. Monitoring of GRO J1655-40, XTE J1550-564, and 4U 1543-47 by RXTE/ASM has shown that the initiation of X-ray outbursts lags behind the start of optical and near-infrared outbursts by 3 to 11 days\citep{orosz_optical_1997,jain_multiwavelength_2001,buxton_2002_2004}. The delays measured with the less sensitive ASM may be overestimated. For GX 339-4, ASM indicated a 20–45 day delay, while PCA showed a delay of less than a week \citep{homan_multiwavelength_2005}.

\subsection{Broadband SED fitting: the contribution of the jet}\label{sec:bbsed}

To better quantify the relative contributions of the disk and jet to OIR emission, broadband SED fitting serves as a useful tool. 
Figure 1 in \cite{gandhi_variable_2011} displays the SED of GX 339-4 during the 2010 outburst around MJD 55266, with the data collected from different wavelength bands covering a time span of up to 10 days. In the MIR band, the spectral index is negative; however, in the OIR regime, it becomes positive. The transition occurs at approximately
$1 \times 10^{14}$ Hz, which can be attributed to a shift in the predominant source of emission from the jet to the disk. \cite{ozbey_arabaci_multiwavelength_2022} and \cite{echiburu-trujillo_chasing_2024} investigated the broadband spectra of MAXI J1820+070, modeling the jet and irradiated disk emission using the \texttt{bnkpower} and \texttt{diskir} models\citep{gierlinski_x-ray_2008,gierlinski_reprocessing_2009}. 
As shown in Figure 2 of their paper, the broadband SEDs indicate that before October 19 (during the decaying hard state), the irradiated disk dominates over the OIR emission (OIR rebrightening peaking around October 14). After October 22, the jet emerges as the dominant contributor in the OIR band. 

Multi-wavelength observations, spanning from radio to X-rays, were available for the 2010-2011 outburst of GX 339-4 in our sample.
The radio data were obtained from the Australia Telescope Compact Array (ATCA) at 5.5 and 9 GHz \citep{corbel_formation_2013,corbel_universal_2013}. Note that, during the decaying phase, only the 9 GHz data were available. We utilized the radio observations as a reference, and the OIR, UV, and X-ray data collected within a $\pm 1$ day window of these observations. There are 31 epochs with simultaneous multiwavelength data: 25 data sets from the rising hard and intermediate states between MJD 55218 and MJD 55315, and 6 data sets from the decaying hard state between MJD 55585 and MJD 55610.

In this study, we employed the same method and parameter constraints previously applied to MAXI J1820+070 \citep{echiburu-trujillo_chasing_2024} to study the OIR origin. 
We employed the model \texttt{\small{redden*tbabs(highecut*bknpower+diskir)}}, while disregarding the emission from the companion star due to its negligible contribution compared to that of the black hole. Furthermore, we used a color excess of \(E_{B-V} = 1.1 \pm 0.2\,\mathrm{mag}\) \citep{buxton_optical_2003}.



\begin{figure}
    \centering  
    \includegraphics[width=0.46\textwidth]{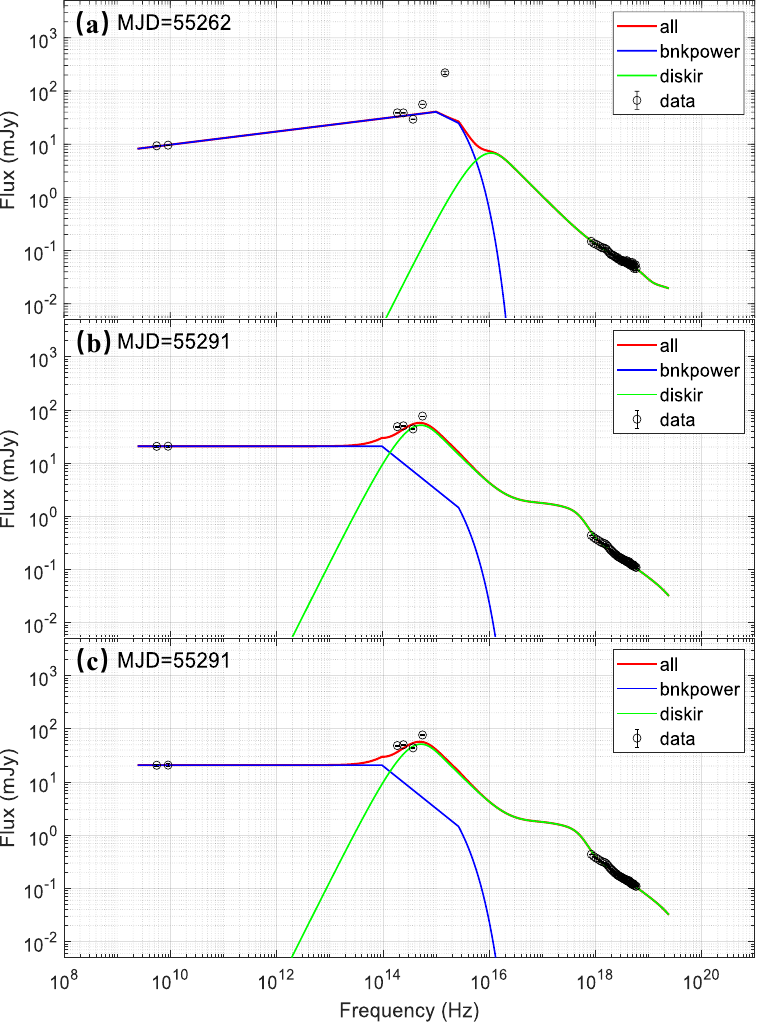}   
    \caption{The broadband spectrum of GX 339-4 during three epochs of its 2010–2011 outburst: panel (a) for MJD 55262, panel (b) for MJD 55291, and panel (c) for MJD 55608.
 The points represent the data, and the red solid line represents the best-fit model. The blue line represents the \texttt{bnkpower} component, and the green line represents the \texttt{diskir} component. The optical, UV, and X-ray data are corrected for reddening and absorption. The number in the top-left corner represents the MJD time. The best broadband-fit parameters are shown in Table \ref{tab:bbsed}.}
    \label{fig:bbsed}
\end{figure}

Regarding the rising hard state, the fitting results suggest that the OIR emission was more likely to originate from a jet before MJD 55272. Figure \ref{fig:bbsed}a illustrates a fit for a specific epoch (MJD 55262), demonstrating the decomposition of the SED for this instance, where the OIR emission is primarily attributed to the jet component modeled by \texttt{bknpower}.
After MJD 55272, the disk emission begins to dominate the OIR band. An example is shown in Figure~\ref{fig:bbsed}b for MJD 55291 when the OIR flux peaks. At this epoch, the disk contributes 67\% of the H-band emission (33\% from the jet), 77\% in the J band (23\% from the jet), 87\% in the I band (13\% from the jet), and 91\% in the V band (9\% from the jet).
It is important to note that, in our fitting, the observed UV flux (\(\nu_{\mathrm{UVW2}} = 1.5 \times 10^{15}\,\mathrm{Hz}\)) significantly exceeds the model predictions (see Fig.~\ref{fig:bbsed}a), likely due to the large extinction correction. The uncertainty in \(E_{B-V}\) has a strong impact on the derived unabsorbed flux; for the adopted value, the extinction in the UVW2 band is \(A_{\mathrm{UVW2}} = 9.6\). Nevertheless, if the UV data are excluded, the SEDs can be fitted better.

During the decaying state, the SED fits indicate that the OIR emission primarily originates from the irradiated disk. However, we cannot constrain the contributions from the jet due to the lack of higher-frequency radio and MIR data. This finding aligns with the conclusions drawn from the time-delay and DIM studies discussed in previous sections. An illustrative example is shown in Figure~\ref{fig:bbsed}c for MJD 55608. 
During this period, the OIR emission is primarily attributed to the accretion disk, with a minor contribution from the jet. Specifically, the disk could account for 92\% of the flux in the H band (with 8\% from the jet), 96\% in the J band (4\% from the jet), 98\% in the I band (2\% from the jet), and 99\% in the V band (1\% from the jet). It is also important to note that the luminosity of the jet is highly uncertain and significantly depends on the model used.



\begin{figure}
    \centering  
    \includegraphics[width=0.46\textwidth]{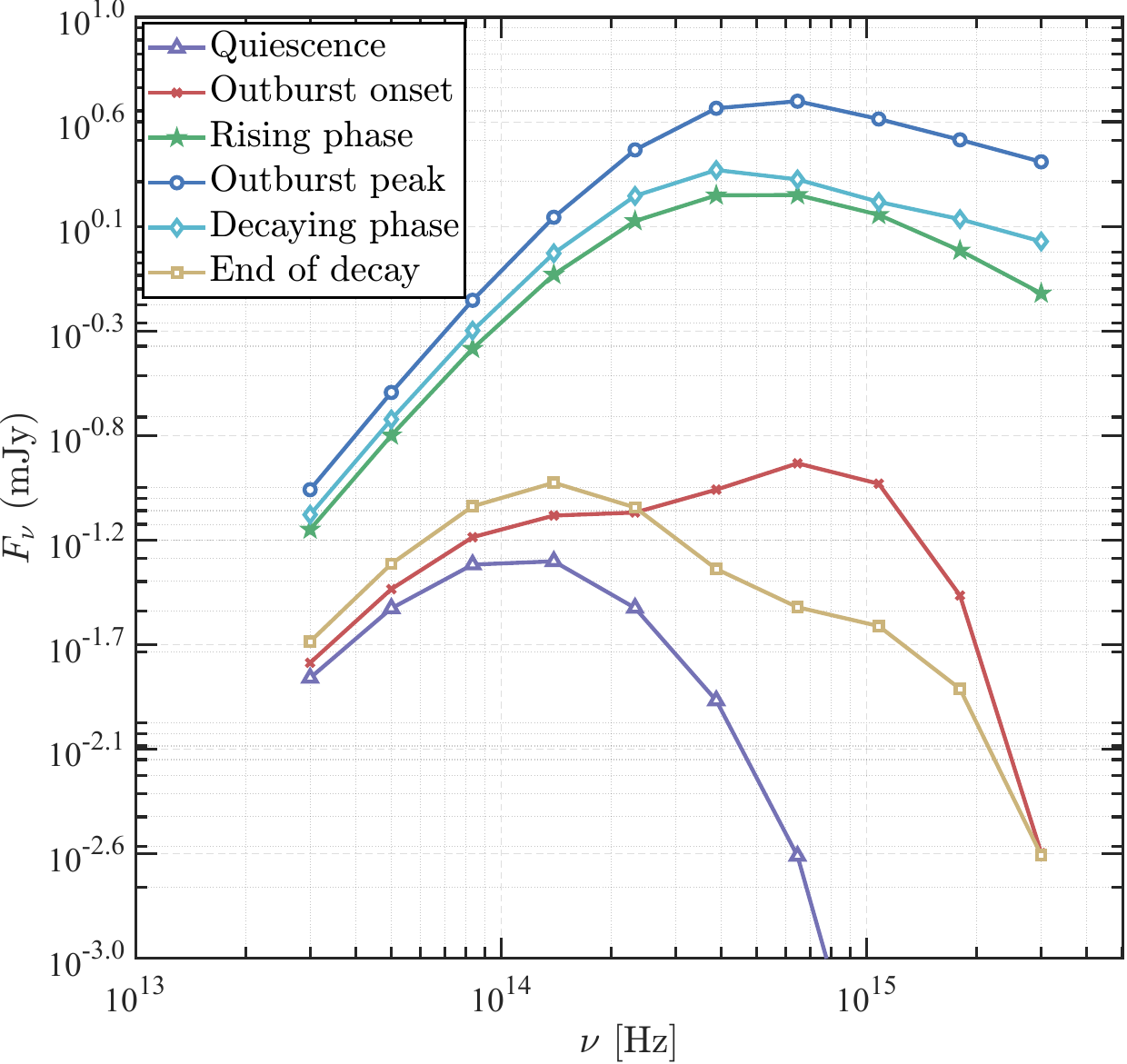}   
    \caption{The DIM-predicted spectrum of a single outburst, using the same input parameters as the GX 339-4 simulation from MJD 52300 to 53000 (see Table \ref{tab:par}).}
    \label{fig:DIMsed}
\end{figure}

In broadband SED modeling, the \texttt{diskir} model provides only steady-state solutions, whereas the DIM offers time-dependent (or time-evolving) solutions for the disk. As illustrated in Fig.~\ref{fig:DIMsed}, the spectra predicted by the DIM show that during quiescence, the flux in the optical to ultraviolet range is very low, with the spectral peak located near the K band. When an outburst begins, the optical to ultraviolet flux rises rapidly due to the quick inward movement of the truncated inner disk radius. From the rising phase to the peak of the outburst, as the accretion rate increases, the high-energy flux grows more rapidly, and the spectral peak gradually shifts to higher frequencies. As the accretion rate starts to decline during the decay phase, this process reverses. Toward the end of the decay, the optical to UV flux decreases more rapidly, resulting in a spectrum that is significantly different from that observed at the onset of the outburst.

One should mention that this work discusses the colour-magnitude diagram, the fast OIR variability and OIR spectroscopy and that these indeed imply the existence of jets which contribute to the OIR, and also mention that this contribution is not the dominant one at the time of reflares, at least in the optical.

\renewcommand\tabcolsep{6pt}
\begin{table*}
\caption{Spectral Fit Parameters for GX 339-4}
\label{tab:bbsed}
\centering
\small
\begin{tabular}{ccccccccccc}
\toprule
 & \multicolumn{4}{c}{\texttt{bknpower}} & \multicolumn{5}{c}{\texttt{diskir}} & \\
\cmidrule(lr){2-5}
\cmidrule(lr){6-10}
MJD & $\Gamma_{\mathrm{1}}$ & $\nu_{\mathrm{b}}$ & $\Gamma_{\mathrm{2}}$ & BPL Norm &
$kT_{\mathrm{disk}}$ & $\Gamma$ & $L_{\mathrm{c}}/L_{\mathrm{d}}$ & $f_{\mathrm{out}}$ & Disk Norm & $\chi^2$ \\
 &  & (Hz) &  &  & (keV) & & & ($\times 10^{-3}$) & ($\times 10^{3}$) & \\
\midrule
55262 & $0.88 \pm 0.001$ & $1 \times 10^{15}$ & $1.5$ & $32.7\pm1.54$ & $3.82 \pm 1.57$ & $1.38 \pm 0.36$ & $10$ & $35.3 \pm 27.7$ & $0.27 \pm 0.37$ & 39.2 \\
55291 & $1.0 \pm 0.3$ & $9.65 \times 10^{13}$ & $1.8$ & $34.51$ & $0.29 \pm 0.05$ & $1.73 \pm 0.005$ & $10$ & $5.40 \pm 0.70$ & $18.57 \pm 6.07$ & 45.70 \\
55608 & $0.90$ & $2.42 \times 10^{11}$ & $1.5$ & $39.33$ & $0.32 \pm 0.07$ & $1.69 \pm 0.025$ & $10$ & $32.81 \pm 4.77$ & $0.85 \pm 0.36$ & 17.50 \\
\bottomrule
\end{tabular}

\vspace{1ex}

\begin{flushleft}
Best broadband-fit parameters of GX 339-4. 
The jet component is modeled with \texttt{bknpower}, where \( \Gamma_1 \) and \( \Gamma_2 \) represent the photon indices of the optically thick and optically thin synchrotron emission, respectively, with BPL Norm as the normalization, and the photon index \( \Gamma \) is related to the spectral index \( \alpha \) via \( \alpha = 1 - \Gamma \).
The accretion flow is modeled with \texttt{diskir}, where $kT_{\mathrm{disk}}$ is the innermost temperature of the unilluminated disk, $\Gamma$ is the photon index of the Comptonized X-ray emission, $L_{\mathrm{c}}/L_{\mathrm{d}}$ is the ratio of the luminosity in the Comptonized emission ($L_{\mathrm{c}}$) to the disk intrinsic luminosity ($L_{\mathrm{d}}$), f$_{\mathrm{out}}$ is the fraction of flux intercepted by the outer disk, and Disk Norm is the disk normalization. $\chi^2$ is the reduced chi-squared of the fit. 
In addition, the high-energy cutoff \( E_{\text{cut}} \) and folding energy \( E_{\text{fold}} \) are fixed at 0.01 keV. The parameters related to the irradiated disk portion of the model, which are fixed across all epochs to typical black hole X-ray binary values, include: the corona temperature \( T_{\mathrm{e}} = 100 \, \mathrm{keV} \), the radius of the illuminated disk \( R_{\mathrm{irr}}/R_{\mathrm{in}} = 1.2 \) (where \( R_{\mathrm{in}} \) is the disk's inner radius), the fraction of luminosity in the Comptonized tail that is thermalized in the inner disk \( f_{\mathrm{in}} = 0.1 \), and the radius of the outer disk, \( R_{\mathrm{out}} = 10^{5} \, R_{\mathrm{in}} \)\citep{echiburu-trujillo_chasing_2024}. Note that parameters without errors cannot be well-constrained. 

\end{flushleft}
\end{table*}

 



\section{Summary}
\label{sec:sum}
We conducted a comprehensive time delay analysis using multi-wavelength data from a sample of well-known BHXRBs. Due to the limited availability of OIR data, our analysis focused on GX 339-4, 4U 1543-47, and XTE J1550-564. We used the ICCF to examine the time delay between Compton luminosity and the OIR band. Additionally, the broadband SED was analyzed to explore the origin of the OIR emission. Our main findings are as follows:

\begin{itemize}
    \item During the rising hard state, the Compton luminosity consistently lags behind OIR emissions by 3-8 days, suggesting that such a delay is a common characteristic in the rising hard state.
    \item  During the rising hard state, the accretion rate at the truncation radius, \(\dot{M}_{\rm{tr}}\), is expected to lag behind the V-band light curve by a few days, according to the DIM. Furthermore, since the Compton luminosity increases with increasing mass accretion rate at the truncation radius (e.g. it goes as \(L_{\rm{C}} \propto \dot{M}_{\rm{tr}}^2\) in the case of the ADAF model), this relationship can account for the observed time delay during the rising hard state. These findings suggest that the OIR emission likely originates from the viscously heated disk during the rising hard state.
    \item In contrast, during the decaying hard state, the OIR emission lags behind the Compton luminosity by approximately 6–35 days. Such a delay is also observed in MAXI J1820+070, indicating that the OIR delay relative to Compton emission is a common feature in the decaying hard state of BHXRBs.
    \item The DIM, when incorporating the effect of the winds, has been successfully applied to explain the delay of optical emissions relative to X-ray fluxes in GX 339-4 and 4U 1543-47 during the decaying hard state.
\end{itemize} 

\section*{Acknowledgements}

B.Y. is supported by the National Program on Key Research and Development Project 2021YFA0718500; by Natural Science Foundation of China (NSFC) grants 12322307, 12361131579, and 12273026;  by Xiaomi Foundation / Xiaomi Young Talents Program. XWC is supported by NSFC (12533005, 12233007, 12347103 and 12361131579), the science research grants from the China Manned Space Project with No. CMS-CSST- 2021-A06, and the fundamental research fund for Chinese central universities (Zhejiang University). This paper has made use of up-to-date SMARTS optical/near-infrared light curves that are available at www.astro.yale.edu/smarts/xrb/home.php. The Yale SMARTS XRB team is supported by NSF grants 0407063 and 070707 to Charles Bailyn. This research has made use of NASA's Astrophysics Data System Bibliographic Services.

\section{DATA AVAILABILITY}
All X-ray data from RXTE are publicly available online. OIR data for XTE J1550-564 and 4U 1543-47 can be found in \cite{jain_multiwavelength_2001, buxton_2002_2004}. OIR data for GX 339-4, obtained from SMARTS, are also publicly available online.




\bibliographystyle{mnras}
\bibliography{ref} 




\appendix

\onecolumn
\section{Multi-wavelength lightcurves and results of ICCF}\label{sec:figlightcurves}
\FloatBarrier
\vspace{-1em}
\begin{figure*}  
    \centering
    \includegraphics[width=0.9\textwidth]{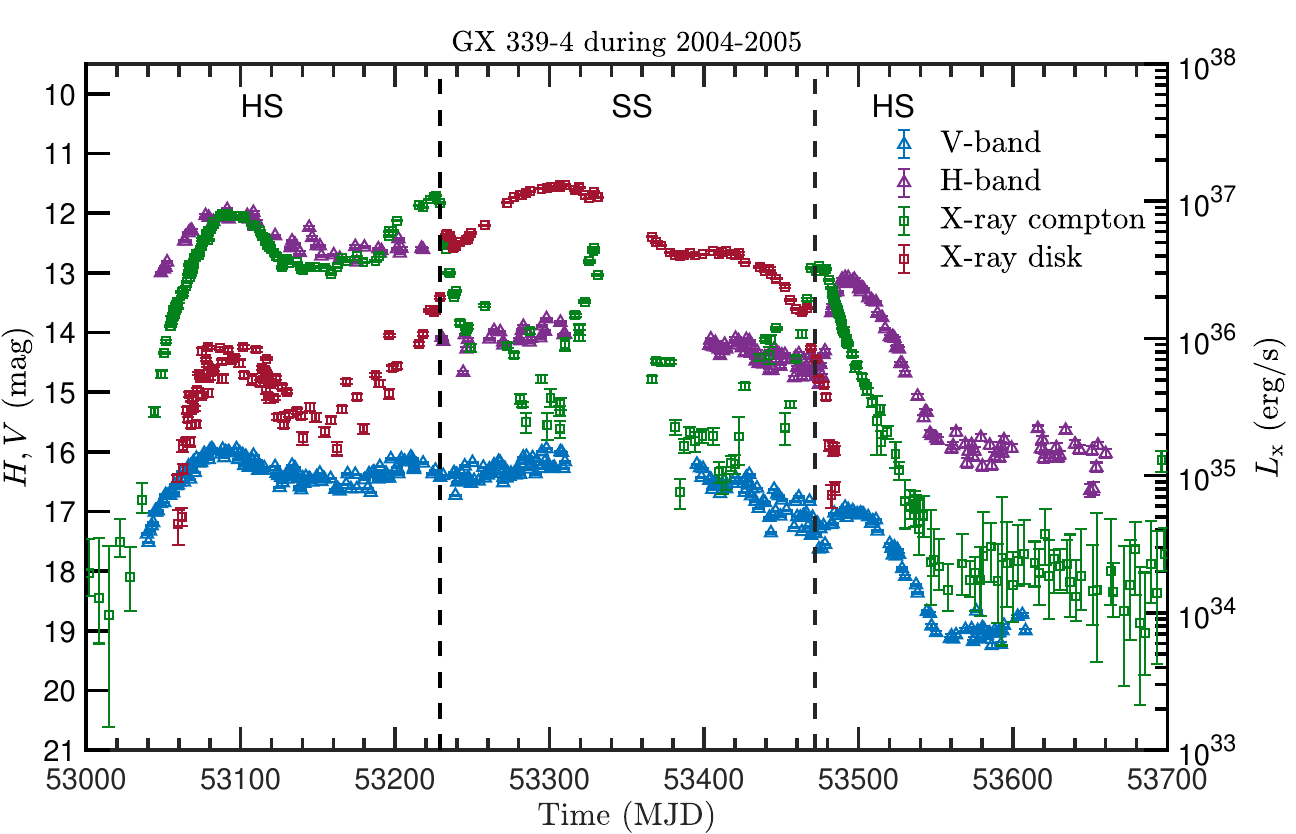}
    \caption{The multi-wavelength lightcurves of GX339-4 from MJD 53000 to 53700. The dashed lines correspond to the Hard state (HS) and Soft state (SS) transitions.}
    \label{fig:339_2}
\end{figure*}
\begin{figure*}
    
    \centering  
    \includegraphics[width=0.8\textwidth]{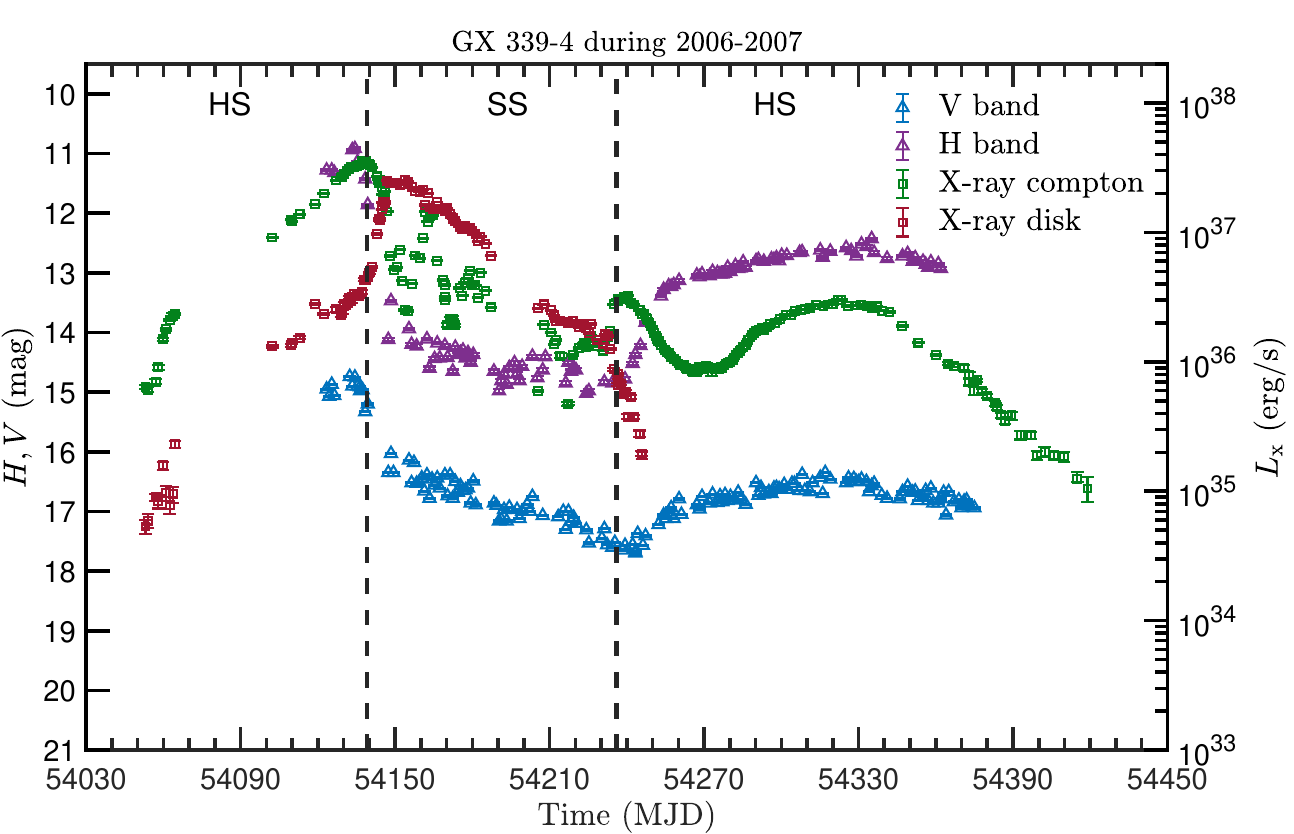}
    \caption{The multi-wavelengths lightcurves of GX339-4 from MJD 54030 to 54450, the dashed lines correspond to the Hard state (HS) and Soft state (SS) transition. 
    \label{fig:339_3}}

    \centering  
    \includegraphics[width=0.8\textwidth]{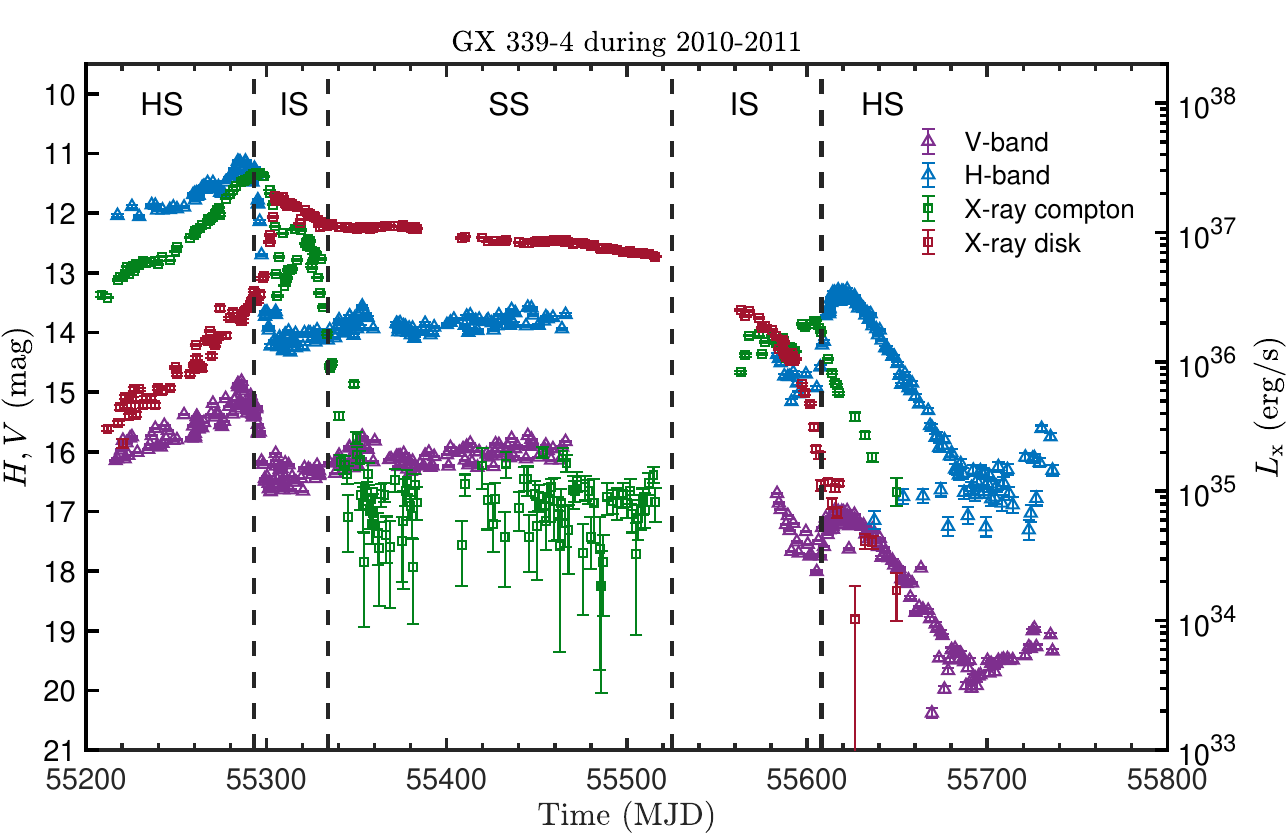}
    \caption{The multi-wavelengths lightcurves of GX339-4 from MJD 55200 to 55800,the dashed lines correspond to the Hard state (HS), Intermediate state (IS), and Soft state (SS) transition.
    \label{fig:339_4}}

\end{figure*} 

\begin{figure*}
    \centering  
    \includegraphics[width=0.8\textwidth]{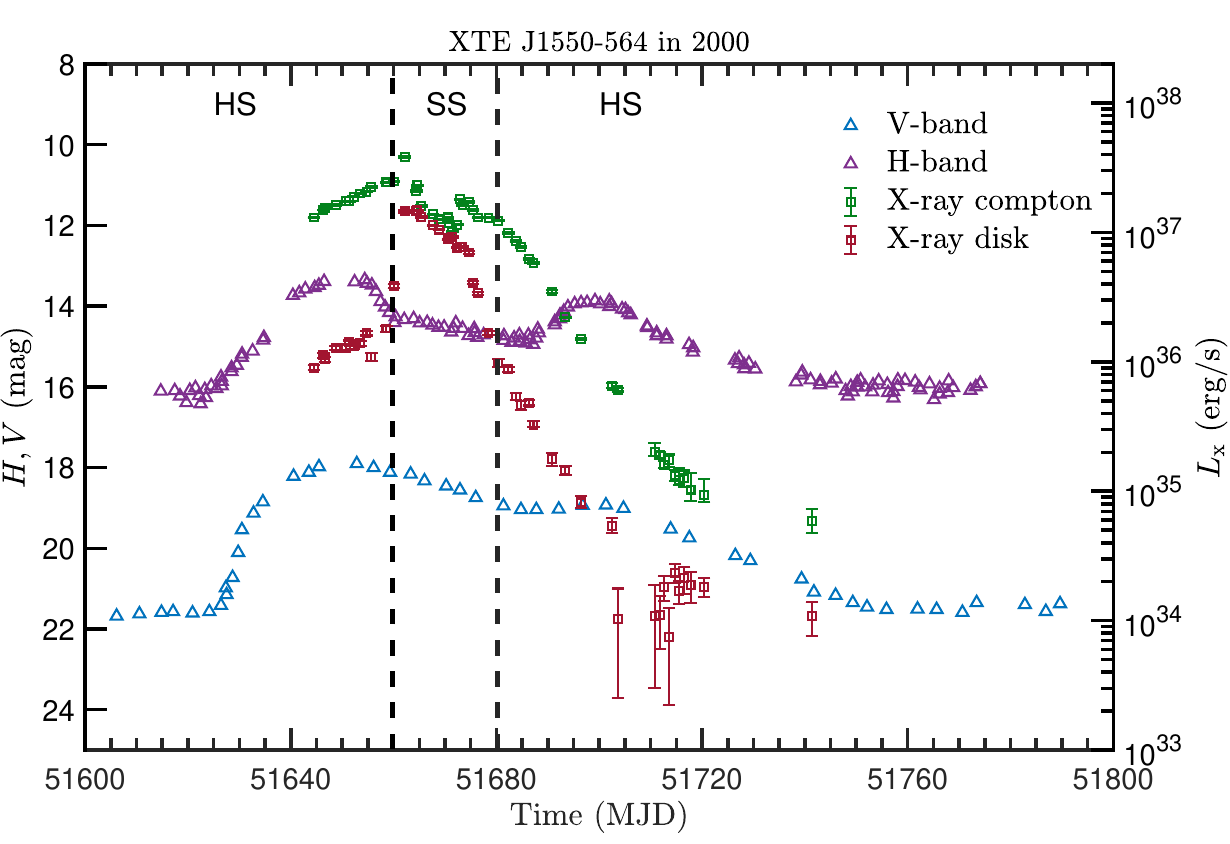}
    \caption{The multi-wavelength lightcurves of XTE J1550-564 from MJD 51600 to 51800. The dashed lines correspond to the Hard state (HS) and Soft state (SS) transitions.}
    \label{fig:1550}
    \centering  
    \includegraphics[width=0.8\textwidth]{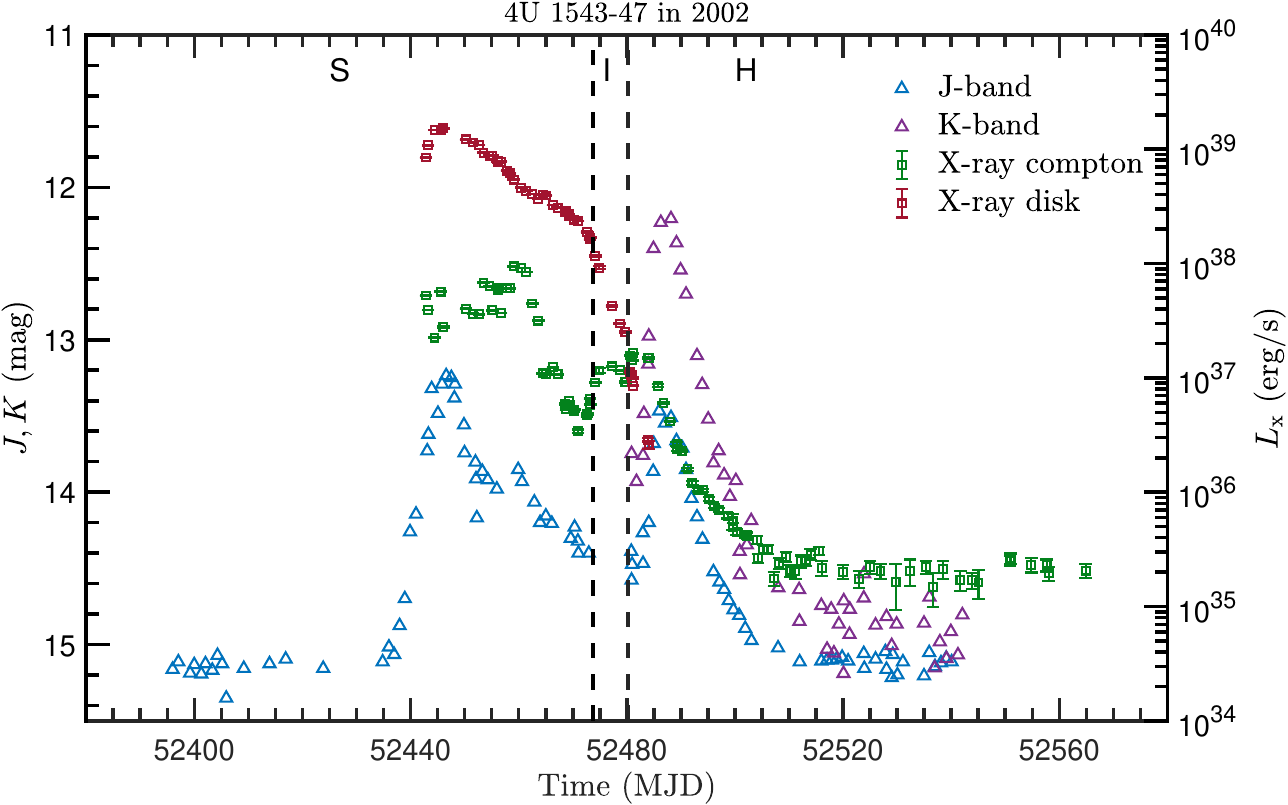}
    \caption{The multi-wavelength lightcurves of 4U 1543-47 from MJD 52380 to 52580. The dashed lines correspond to the Hard state (HS), Intermediate state (IS), and Soft state (SS) transitions.}
    \label{fig:1543}
\end{figure*}

\begin{figure}
	\centering

	\includegraphics[width=\textwidth]{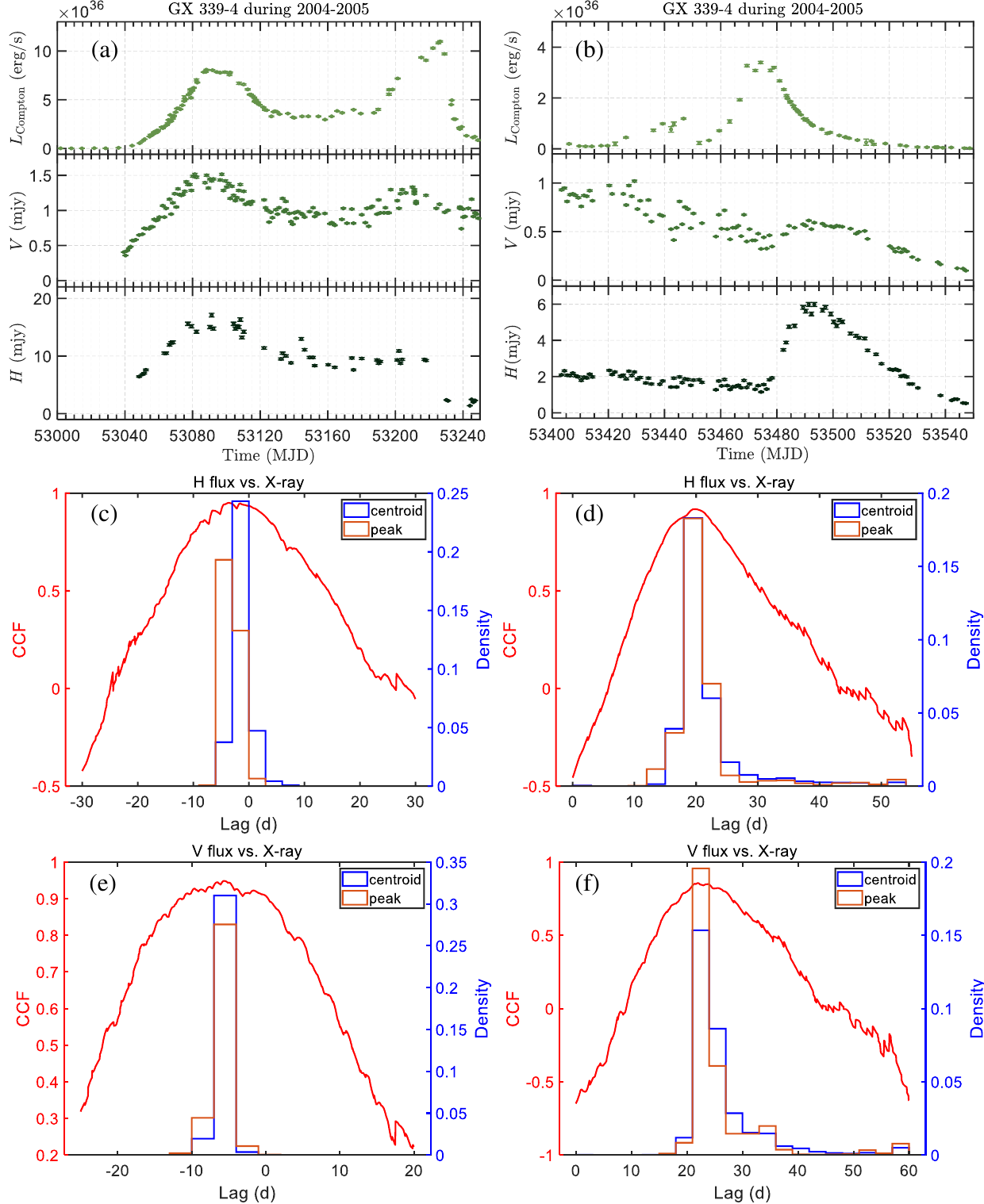}  
	\caption{Same as Fig. \ref{fig:i339_1} for the 2002-2003 outburst of GX 339-4}
        
	\label{fig:i339_2}
\end{figure}

\begin{figure}
	\centering
        
	\includegraphics[width=\textwidth]{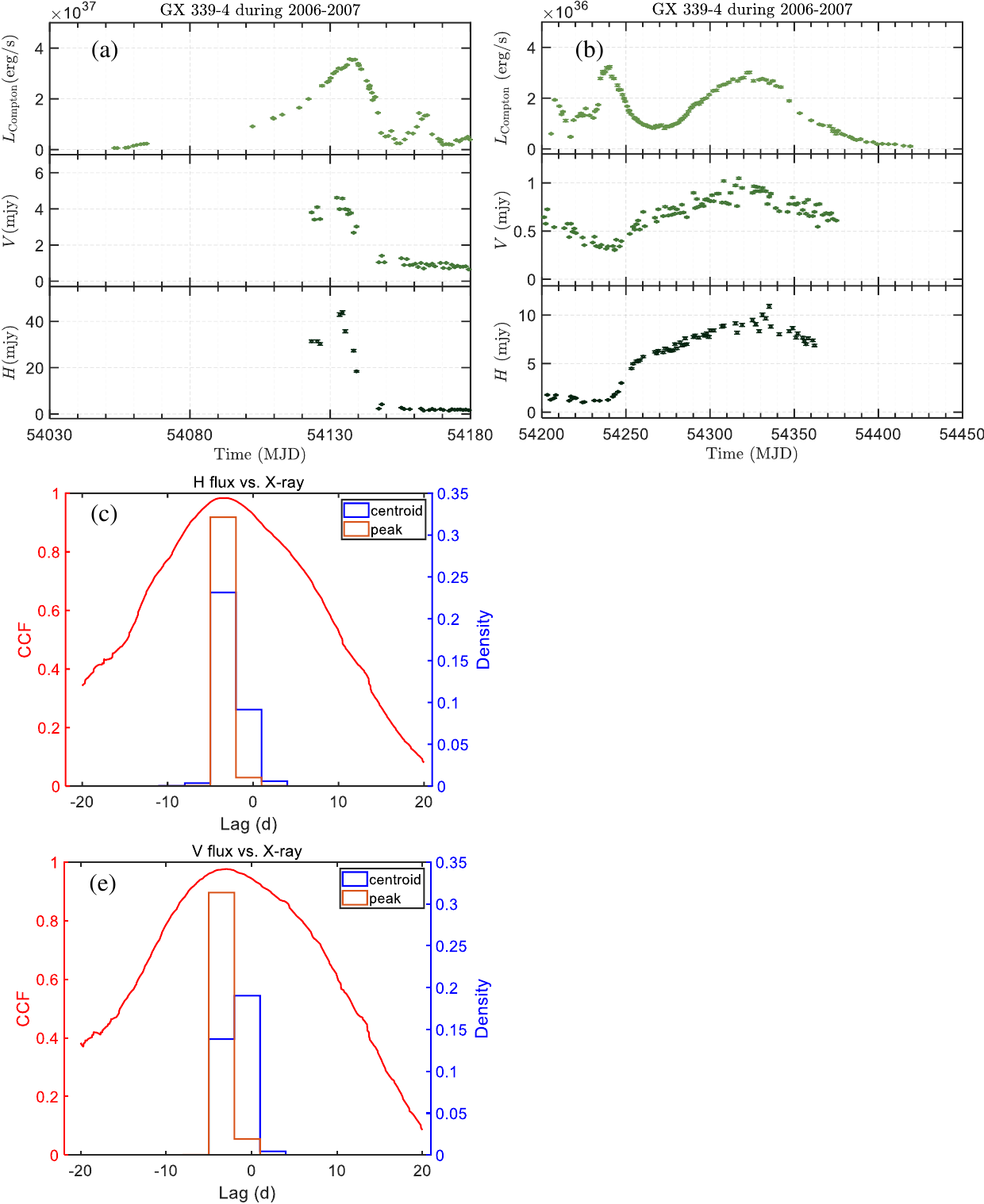}  
	\caption{Same as Fig. \ref{fig:i339_1} for the 2006-2007 outburst of GX 339-4, the ICCF analysis has not been performed in H,V during decaying (see text).}
        
	\label{fig:i339_3}
\end{figure}

\begin{figure}
	\centering
        
	\includegraphics[width=\textwidth]{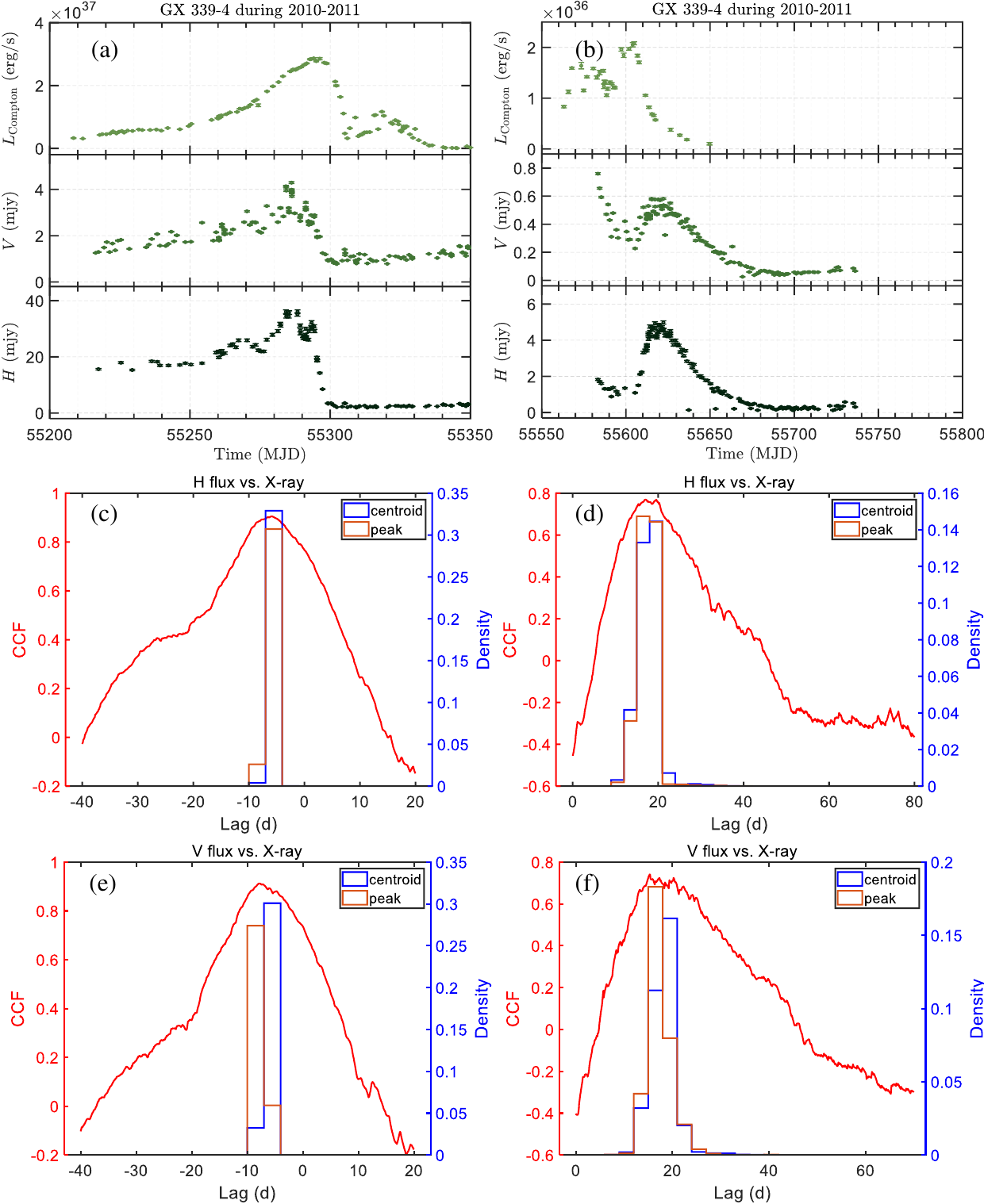}  
	\caption{Same as Fig. \ref{fig:i339_1} for the 2010-2011 outburst of GX 339-4}
        
	\label{fig:i339_4}
\end{figure}

\begin{figure}
	\centering
        
	\includegraphics[width=\textwidth]{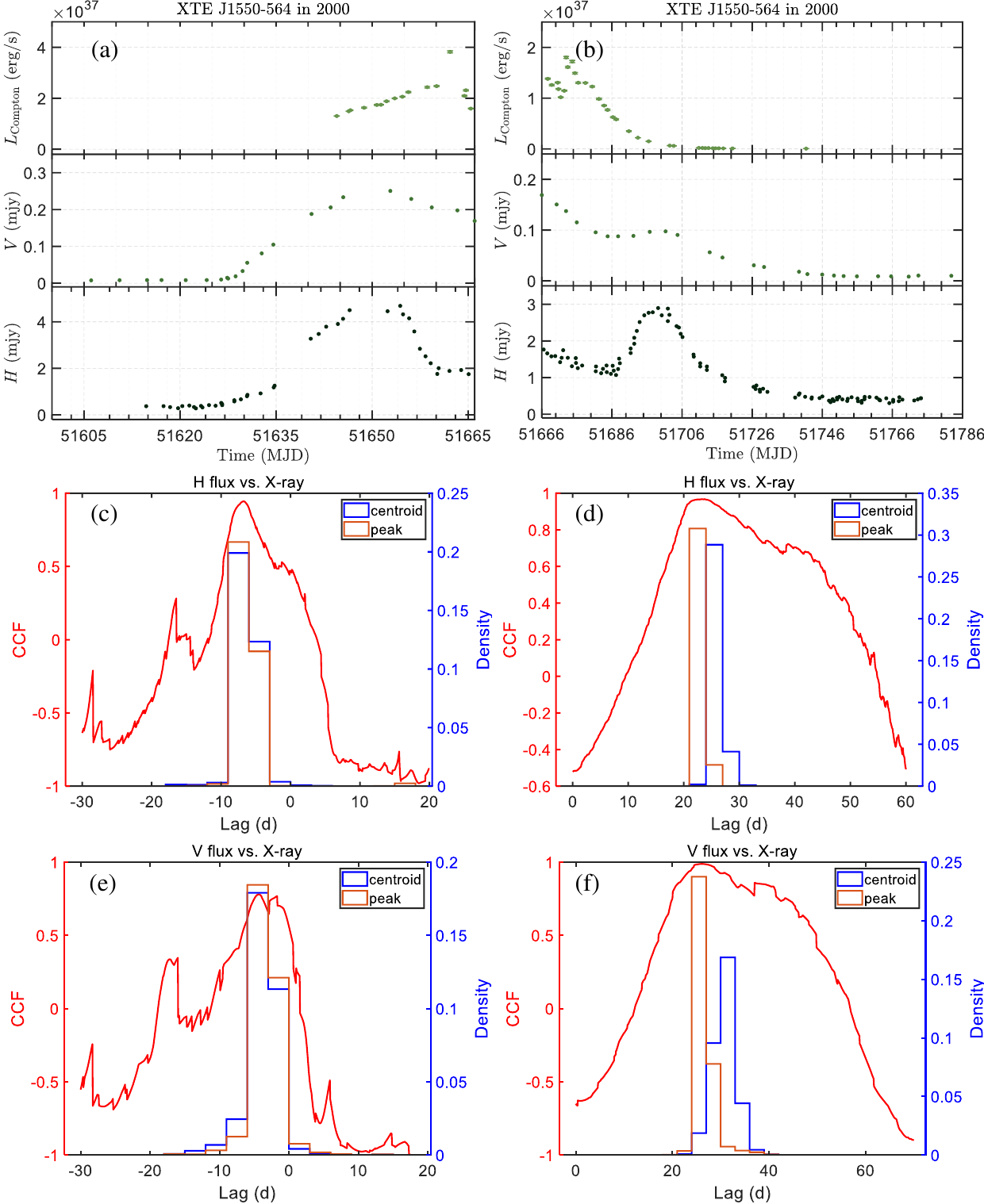}  
	\caption{Same as Fig. \ref{fig:i339_1} for XTE J1550-564}
        
	\label{fig:i1550}
\end{figure}

\begin{figure}
	\includegraphics[width=\textwidth]{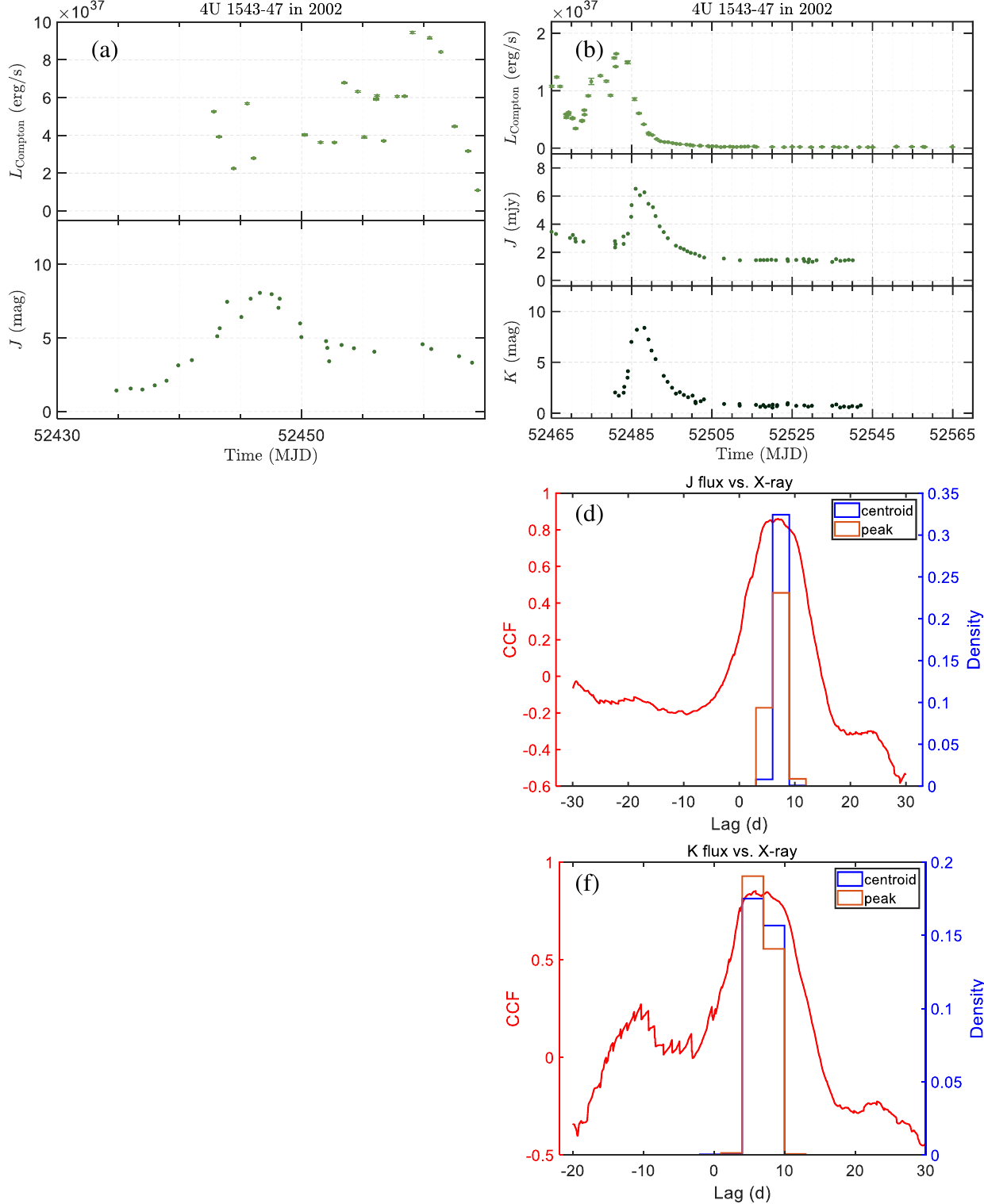}

	\caption{Same as Fig. \ref{fig:i339_1} for 4U 1543-47. J and K bands are displayed instead of H and V bands, and as no K data is not available during rise, the ICCF analysis has not been performed in J,K during rise (see text).}
        
	\label{fig:i1543}
\end{figure}

\FloatBarrier
\section{Decaying hard state DIM simulation }\label{sec:figwind}
\FloatBarrier
\begin{figure}  
    \centering  
    \includegraphics[width=0.8\columnwidth]{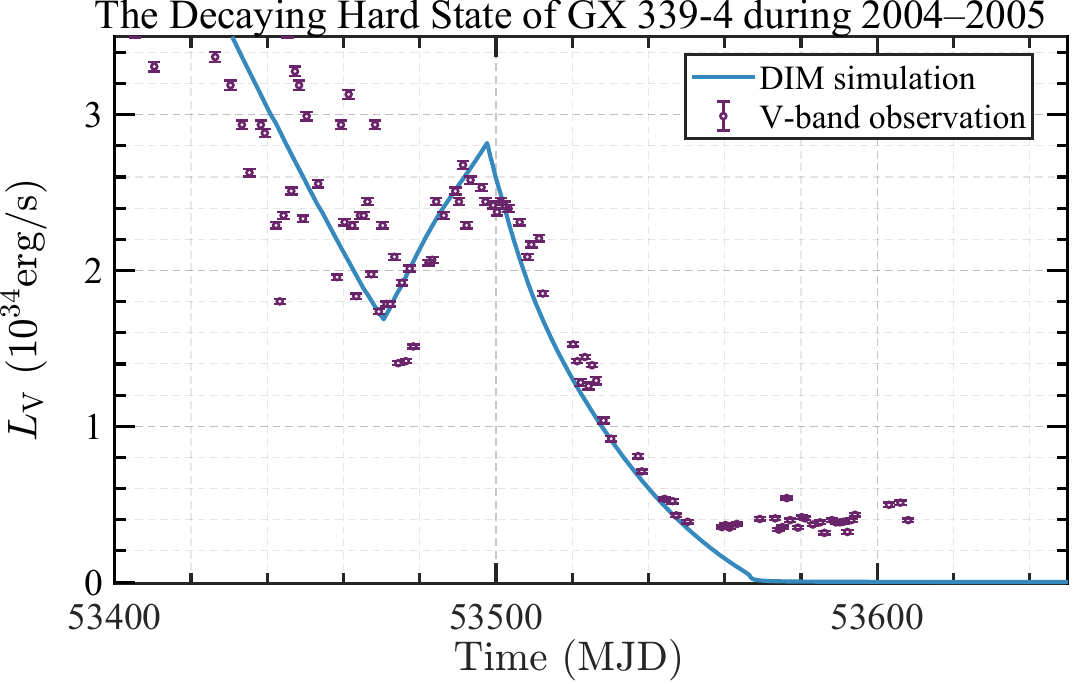}  
    \caption{Disk instability model with the disk wind simulation of the 2004-2005 outburst of GX 339-4. The blue line represents the simulation results, while the purple dots indicate the observed data.}  
    \label{fig:339s2}  
\end{figure}  
\begin{figure}
    \centering  
    \includegraphics[width=0.8\columnwidth]{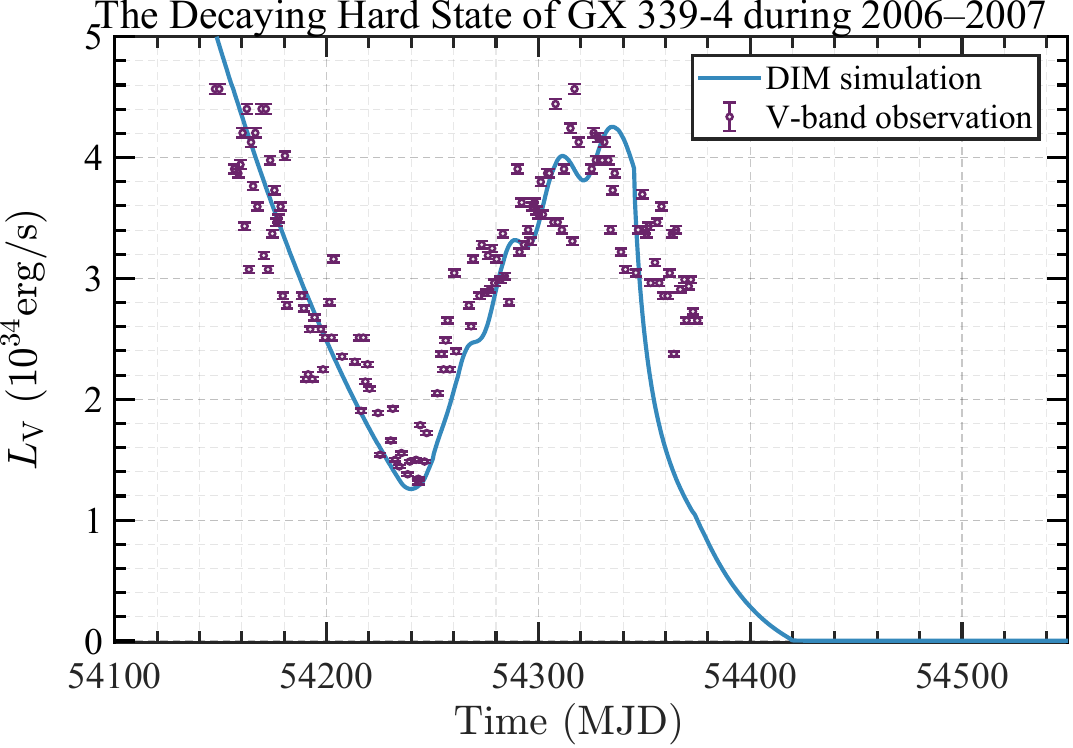}   
    \caption{Disk instability model with the disk wind simulation of the 2006-2007 outburst of GX 339-4. The blue line represents the simulation results, while the purple dots indicate the observed data.}
    \label{fig:339s3}
\end{figure}

\begin{figure}
    \centering  
    \includegraphics[width=0.8\columnwidth]{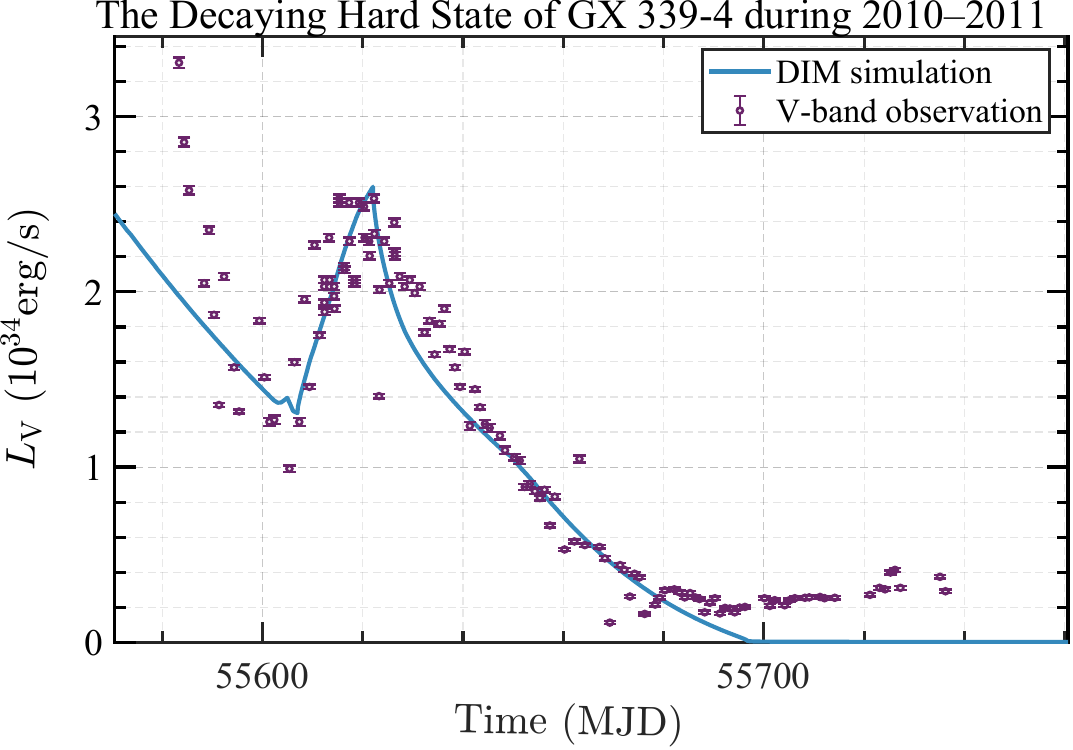}   
    \caption{Disk instability model with the disk wind simulation of the 2010-2011 outburst of GX 339-4. The blue line represents the simulation results, while the purple dots indicate the observed data.}  
    \label{fig:339s4}  
\end{figure}  

\begin{figure}
    \centering  
    \includegraphics[width=0.8\columnwidth]{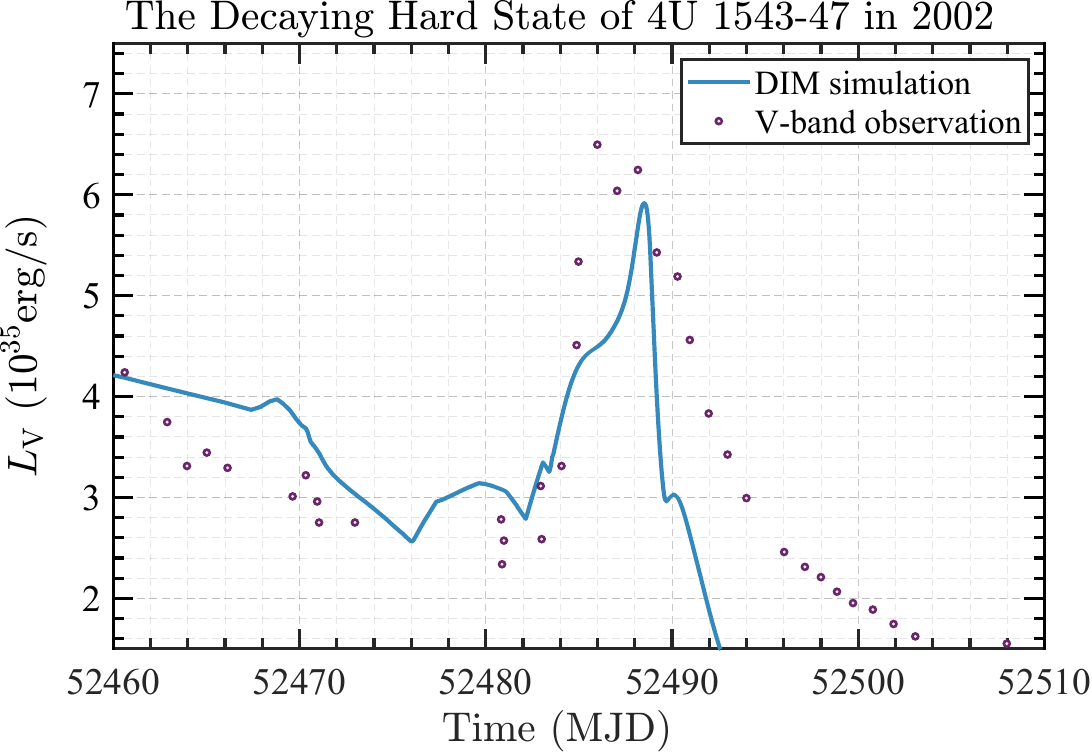}   
    \caption{Disk instability model with the disk wind simulation of 4U 1543-47 in 2002. The blue line represents the simulation results, while the purple dots indicate the observed data.}  
    \label{fig:1543s}  
\end{figure} 

\begin{figure}
    \centering  
    \includegraphics[width=0.75\columnwidth]{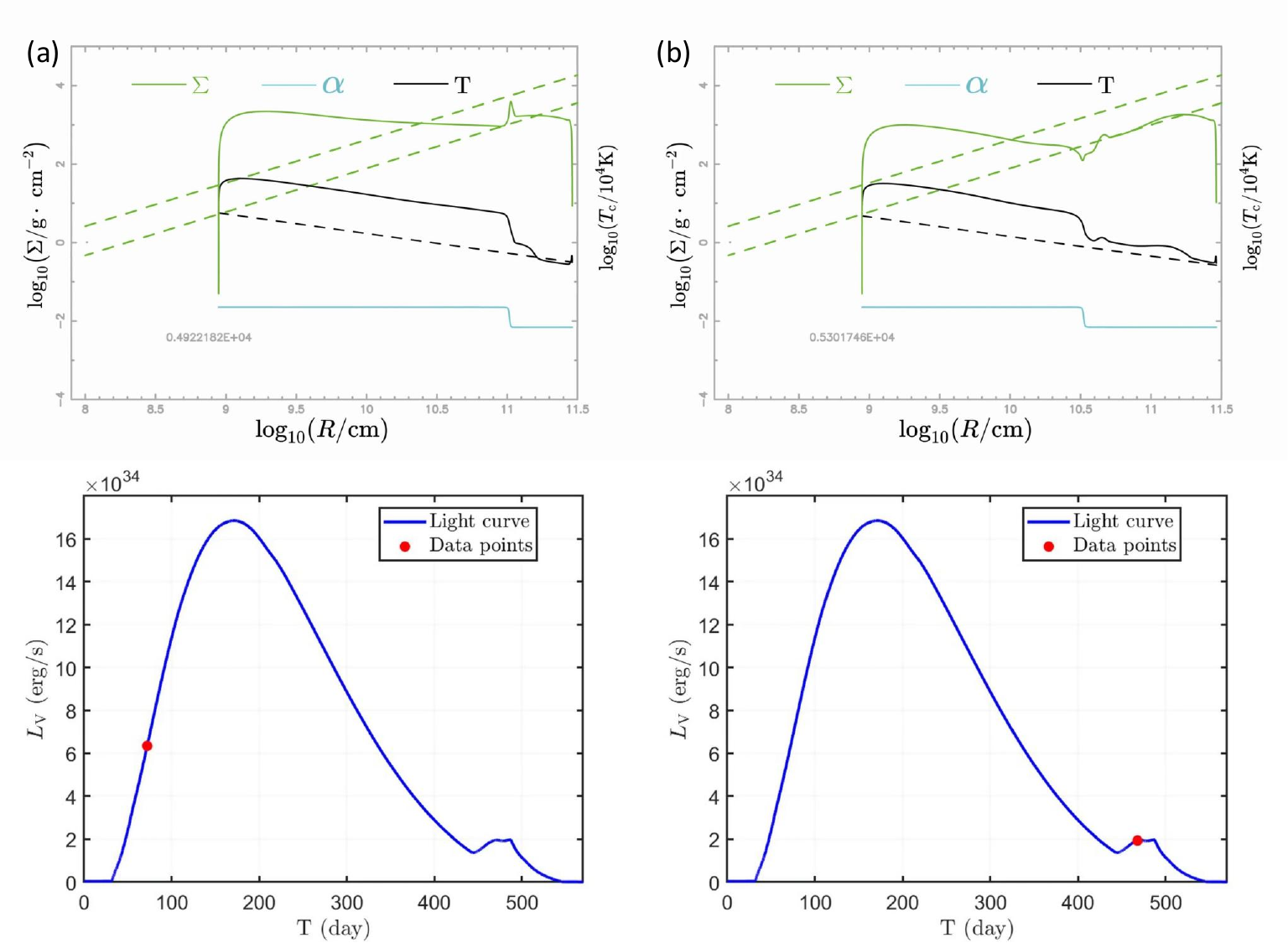}   
    \caption{The other two frames of the animation in Fig.~\ref{fig:animation_339_2002} are shown here. Panel a corresponds to the 120th frame of the animation, and panel b corresponds to the 200th frame.} 
    \label{fig:DIM_frame}  
\end{figure} 
\FloatBarrier

\section{Spectral Fitting tables and figure}\label{sec:spectral}
\vspace{-20cm}
\renewcommand{\arraystretch}{1.3}
\begin{table*}
\centering
\caption{Spectral Fit Parameters for GX 339--4}
\label{tab:339spectral}
\begin{tabular}{ccccccccccc}
\hline
MJD & $T_{\rm in}$ & $N_{\rm disk}$ & $\sigma$ & $N_{\rm gauss}$ &
$\Gamma$ & $kT_{\rm bb}$ & $N_{\rm nth}$ & $\chi^2_{\nu}$ &
$L_{\rm Disk}$ & $L_{\rm Comp}$ \\
(d) & (keV) & ($10^{-2}$) & (keV) & ($10^{-3}$) &
 & (keV) & ($10^{-2}$) & &
($10^{36}\,\mathrm{erg\,s^{-1}}$) & ($10^{36}\,\mathrm{erg\,s^{-1}}$) \\
\hline
52372 & $1.19^{+0.13}_{-0.22}$ & $0.50^{+0.53}_{-0.14}$ & $0.81^{+0.33}_{-0.17}$ & $8.88^{+5.40}_{-2.18}$ & $1.54^{+0.03}_{-0.03}$ & $1.19^{+0.13}_{-0.22}$ & $5.83^{+3.07}_{-1.26}$ & 0.66 & $2.22^{+0.06}_{-0.06}$ & $18.53^{+0.14}_{-0.13}$ \\
52373 & $1.03^{+0.14}_{-0.13}$ & $1.04^{+0.71}_{-0.37}$ & $0.96^{+0.30}_{-0.25}$ & $15.24^{+5.68}_{-4.34}$ & $1.58^{+0.02}_{-0.03}$ & $1.03^{+0.14}_{-0.13}$ & $10.08^{+3.09}_{-2.47}$ & 0.82 & $2.08^{+0.08}_{-0.08}$ & $23.65^{+0.18}_{-0.17}$ \\
52374 & $0.95^{+0.11}_{-0.19}$ & $1.45^{+2.32}_{-0.48}$ & $1.09^{+0.21}_{-0.20}$ & $21.05^{+6.85}_{-4.21}$ & $1.58^{+0.03}_{-0.03}$ & $0.95^{+0.11}_{-0.19}$ & $12.35^{+6.43}_{-2.51}$ & 0.75 & $1.77^{+0.08}_{-0.08}$ & $24.83^{+0.19}_{-0.19}$ \\
52377 & $1.01^{+0.07}_{-0.13}$ & $1.41^{+1.03}_{-0.32}$ & $1.03^{+0.18}_{-0.17}$ & $21.89^{+5.92}_{-4.42}$ & $1.59^{+0.02}_{-0.02}$ & $1.01^{+0.07}_{-0.13}$ & $12.65^{+4.28}_{-1.77}$ & 0.71 & $2.42^{+0.09}_{-0.09}$ & $27.70^{+0.19}_{-0.19}$ \\
52379 & $0.96^{+0.11}_{-0.14}$ & $1.72^{+1.46}_{-0.54}$ & $0.87^{+0.21}_{-0.18}$ & $21.44^{+6.48}_{-4.98}$ & $1.62^{+0.02}_{-0.02}$ & $0.96^{+0.11}_{-0.14}$ & $15.92^{+5.94}_{-3.21}$ & 0.81 & $2.24^{+0.10}_{-0.10}$ & $31.06^{+0.22}_{-0.22}$ \\
52381 & $1.01^{+0.11}_{-0.15}$ & $1.76^{+1.66}_{-0.55}$ & $1.16^{+0.15}_{-0.18}$ & $31.98^{+7.30}_{-6.60}$ & $1.60^{+0.03}_{-0.02}$ & $1.01^{+0.11}_{-0.15}$ & $15.00^{+6.08}_{-2.84}$ & 0.82 & $3.04^{+0.10}_{-0.10}$ & $32.61^{+0.22}_{-0.22}$ \\
52382 & $0.98^{+0.09}_{-0.17}$ & $1.87^{+2.28}_{-0.51}$ & $1.10^{+0.14}_{-0.15}$ & $31.42^{+7.02}_{-5.53}$ & $1.62^{+0.02}_{-0.02}$ & $0.98^{+0.09}_{-0.17}$ & $16.47^{+8.07}_{-2.66}$ & 0.54 & $2.79^{+0.10}_{-0.10}$ & $33.57^{+0.22}_{-0.22}$ \\
52383 & $1.00^{+0.11}_{-0.15}$ & $1.69^{+1.36}_{-0.50}$ & $0.96^{+0.19}_{-0.18}$ & $24.56^{+7.28}_{-5.67}$ & $1.64^{+0.02}_{-0.02}$ & $1.00^{+0.11}_{-0.15}$ & $16.97^{+6.81}_{-3.33}$ & 0.67 & $2.76^{+0.11}_{-0.11}$ & $34.61^{+0.23}_{-0.23}$ \\
52384 & $1.01^{+0.13}_{-0.16}$ & $1.75^{+1.62}_{-0.61}$ & $0.99^{+0.16}_{-0.14}$ & $27.78^{+6.48}_{-6.31}$ & $1.64^{+0.02}_{-0.02}$ & $1.01^{+0.13}_{-0.16}$ & $16.83^{+7.24}_{-3.74}$ & 0.73 & $2.97^{+0.10}_{-0.10}$ & $34.94^{+0.22}_{-0.22}$ \\
52385 & $0.98^{+0.07}_{-0.12}$ & $2.02^{+1.29}_{-0.46}$ & $1.11^{+0.13}_{-0.13}$ & $33.63^{+6.00}_{-5.09}$ & $1.62^{+0.02}_{-0.01}$ & $0.98^{+0.07}_{-0.12}$ & $17.19^{+5.03}_{-2.32}$ & 0.67 & $3.00^{+0.10}_{-0.10}$ & $34.78^{+0.19}_{-0.19}$ \\
52385 & $1.05^{+0.12}_{-0.16}$ & $1.54^{+1.44}_{-0.48}$ & $1.04^{+0.19}_{-0.20}$ & $28.48^{+8.83}_{-6.88}$ & $1.63^{+0.02}_{-0.02}$ & $1.05^{+0.12}_{-0.16}$ & $15.42^{+6.35}_{-3.19}$ & 0.55 & $3.30^{+0.11}_{-0.11}$ & $34.96^{+0.24}_{-0.24}$ \\
\hline
\end{tabular}
\begin{flushleft}
MJD: Modified Julian Date. 
$T_{\rm in}$: inner disk temperature from \texttt{diskbb}. 
$N_{\rm disk}$: normalization of the \texttt{diskbb} component. 
$\sigma$: line width of the Gaussian component. 
$N_{\rm gauss}$: normalization of the Gaussian component. 
$\Gamma$: photon index of the \texttt{nthComp} component.  
$kT_{\rm bb}$: seed photon temperature of \texttt{nthComp} (linked to $T_{\rm in}$). 
$N_{\rm nth}$: normalization of the \texttt{nthComp} component. 
$\chi^2$: reduced chi-squared of the fit. 
$L_{\rm Disk}$ and $L_{\rm Comp}$: unabsorbed luminosities in 3–25 keV. 
See Section~\ref{sec:fitting} for details. \\
Table~\ref{tab:339spectral} is published in its entirety in the electronic version of the journal. A portion is shown here for guidance.
\end{flushleft}
\end{table*}

\begin{table*}
\caption{Spectral fit parameters for XTE J1550-564.}
\label{tab:1550spectral}
\centering
\begin{tabular}{ccccccccccc}
\hline
MJD (d) & $Max_\mathrm{tau}$ & $Width$ & $T_\mathrm{in}$  & $N_\mathrm{disk}$  & $\Gamma$ & $N_\mathrm{nth}$  & $N_\mathrm{gauss}$  & $\chi^2$ & $L_\mathrm{D}$  & $L_\mathrm{C}$  \\
(d) &  &  & (keV) & ($10^{2}$) &
 & ($10^{-2}$)& ($10^{-2}$) & &
($10^{36}\,\mathrm{erg\,s^{-1}}$) & ($10^{36}\,\mathrm{erg\,s^{-1}}$) \\ 
\hline
51644 & $0.00^{+\mathrm{NaN}}_{-\mathrm{NaN}}$ & $98.54^{+\mathrm{NaN}}_{-\mathrm{NaN}}$ & $0.85^{+0.04}_{-0.16}$ & $1.90^{+2.56}_{-0.41}$ & $1.54^{+0.03}_{-0.01}$ & $9.35^{+4.81}_{-0.63}$ & $0.88^{+0.11}_{-0.33}$ & 0.61 & $0.93^{+0.04}_{-0.04}$ & $13.00^{+0.06}_{-0.06}$ \\
51646 & $0.00^{+\mathrm{NaN}}_{-\mathrm{NaN}}$ & $99.86^{+\mathrm{NaN}}_{-\mathrm{NaN}}$ & $0.87^{+0.01}_{-0.21}$ & $2.01^{+3.97}_{-0.16}$ & $1.55^{+0.06}_{-0.01}$ & $10.46^{+8.79}_{-0.36}$ & $1.04^{+0.20}_{-0.89}$ & 0.54 & $1.12^{+0.05}_{-0.04}$ & $14.92^{+0.13}_{-0.11}$ \\
51647 & $0.35^{+\mathrm{NaN}}_{-0.25}$ & $31.53^{+65.87}_{-8.96}$ & $0.83^{+0.05}_{-0.12}$ & $2.39^{+2.43}_{-0.55}$ & $1.55^{+0.02}_{-0.01}$ & $11.45^{+4.23}_{-1.08}$ & $1.04^{+0.21}_{-0.49}$ & 0.65 & $1.05^{+0.05}_{-0.05}$ & $15.31^{+0.16}_{-0.16}$ \\
51649 & $0.42^{+2.50}_{-0.33}$ & $90.17^{+5.70}_{-76.51}$ & $0.88^{+0.04}_{-0.10}$ & $2.06^{+1.48}_{-0.45}$ & $1.54^{+0.02}_{-0.01}$ & $10.92^{+3.38}_{-0.77}$ & $1.19^{+0.18}_{-0.58}$ & 0.81 & $1.29^{+0.05}_{-0.05}$ & $16.27^{+0.27}_{-0.15}$ \\
51651 & $0.20^{+\mathrm{NaN}}_{-\mathrm{NaN}}$ & $11.84^{+85.87}_{-\mathrm{NaN}}$ & $0.86^{+0.04}_{-0.11}$ & $2.36^{+2.12}_{-0.49}$ & $1.55^{+0.02}_{-0.02}$ & $12.25^{+3.91}_{-1.19}$ & $1.18^{+0.27}_{-0.56}$ & 0.80 & $1.27^{+0.05}_{-0.05}$ & $17.39^{+0.22}_{-0.19}$ \\
51651 & $0.02^{+\mathrm{NaN}}_{-\mathrm{NaN}}$ & $5.45^{+\mathrm{NaN}}_{-\mathrm{NaN}}$ & $0.88^{+0.04}_{-0.11}$ & $2.27^{+1.92}_{-0.45}$ & $1.55^{+0.02}_{-0.01}$ & $11.85^{+3.96}_{-0.84}$ & $1.34^{+0.18}_{-0.58}$ & 0.56 & $1.41^{+0.05}_{-0.05}$ & $17.44^{+0.19}_{-0.13}$ \\
51652 & $0.26^{+\mathrm{NaN}}_{-0.15}$ & $28.08^{+69.10}_{-5.43}$ & $0.82^{+0.03}_{-0.10}$ & $3.19^{+2.35}_{-0.61}$ & $1.56^{+0.02}_{-0.01}$ & $14.60^{+4.21}_{-1.03}$ & $1.41^{+0.24}_{-0.63}$ & 0.69 & $1.32^{+0.06}_{-0.06}$ & $18.83^{+0.22}_{-0.19}$ \\
51654 & $0.34^{+2.06}_{-0.29}$ & $91.38^{+6.05}_{-72.12}$ & $0.81^{+0.07}_{-0.09}$ & $3.75^{+3.03}_{-1.21}$ & $1.58^{+0.02}_{-0.01}$ & $16.33^{+4.61}_{-2.29}$ & $1.62^{+0.20}_{-0.60}$ & 0.52 & $1.39^{+0.06}_{-0.06}$ & $19.96^{+0.19}_{-0.15}$ \\
51655 & $0.19^{+\mathrm{NaN}}_{-0.12}$ & $98.15^{+\mathrm{NaN}}_{-75.00}$ & $0.85^{+0.05}_{-0.09}$ & $3.23^{+2.07}_{-0.84}$ & $1.57^{+0.02}_{-0.01}$ & $15.33^{+4.02}_{-1.60}$ & $1.78^{+0.18}_{-0.63}$ & 0.48 & $1.64^{+0.06}_{-0.06}$ & $20.63^{+0.18}_{-0.14}$ \\
51656 & $0.35^{+\mathrm{NaN}}_{-0.07}$ & $3.12^{+\mathrm{NaN}}_{-0.72}$ & $0.55^{+0.16}_{-0.09}$ & $20.40^{+50.94}_{-13.68}$ & $1.65^{+0.01}_{-0.06}$ & $40.08^{+12.10}_{-16.40}$ & $0.20^{+1.14}_{-0.14}$ & 0.79 & $0.51^{+0.08}_{-0.08}$ & $23.46^{+0.24}_{-0.22}$ \\
\hline
\end{tabular}
\vspace{0.2cm}
\begin{minipage}{0.97\textwidth}
\vspace{0.2cm}
\begin{flushleft}
MJD: Modified Julian Date. 
$Max_\mathrm{tau}$: the maximum absorption factor at threshold of \texttt{smedge}.
$Width$: Smearing width of \texttt{smedge}.
$T_{\rm in}$: inner disk temperature from \texttt{diskbb}. 
$N_{\rm disk}$: normalization of the \texttt{diskbb} component. 
$\Gamma$: photon index of the \texttt{nthComp} component.
$N_{\rm nth}$: normalization of the \texttt{nthComp} component. 
$N_{\rm gauss}$: normalization of the Gaussian component. 
$\chi^2$: reduced chi-squared of the fit. 
$L_{\rm Disk}$: unabsorbed disk luminosity in $3$–$25$ keV. 
$L_{\rm Compton}$: unabsorbed Comptonized luminosity in $3$–$25$ keV. 
Other details can be found in Sec.\ref{sec:fitting}. 

Table~\ref{tab:1550spectral} is published in its entirety in the electronic version of the journal. A portion is shown here for guidance.
\end{flushleft}
\end{minipage}
\end{table*}

\begin{table*}
\centering
\caption{Spectral fit parameters for 4U 1543$-$47.}
\label{tab:1543spectral}
\begin{tabular}{cccccccccc}
\hline
MJD & $T_{\rm in}$ & $N_{\rm disk}$ & $\sigma$ & $N_{\rm gauss}$ & $\Gamma$ & $N_{\rm nth}$ & $\chi^2$ & $L_{\rm Disk}$ & $L_{\rm Comp}$ \\
(d) & (keV) & ($10^{2}$) & (keV) & ($10^{-3}$) &  & ($10^{-2}$) &  & ($10^{36}\,\mathrm{erg\,s^{-1}}$) & ($10^{36}\,\mathrm{erg\,s^{-1}}$) \\
\hline
52443 & $0.88^{+0.01}_{-0.01}$ & $74.20^{+4.39}_{-3.80}$ & $0.52^{+0.18}_{-0.22}$ & $14.04^{+4.97}_{-4.43}$ & $2.28^{+0.03}_{-0.03}$ & $13.25^{+0.81}_{-0.75}$ & 0.47 & $844.98^{+4.96}_{-4.97}$ & $52.53^{+0.27}_{-0.27}$ \\
52443 & $0.89^{+0.01}_{-0.01}$ & $90.69^{+4.15}_{-3.77}$ & $0.42^{+0.13}_{-0.16}$ & $19.32^{+4.83}_{-4.51}$ & $2.35^{+0.05}_{-0.05}$ & $10.36^{+0.84}_{-0.77}$ & 0.66 & $1082.17^{+5.90}_{-5.88}$ & $39.26^{+0.29}_{-0.29}$ \\
52444 & $0.95^{+0.00}_{-0.01}$ & $96.74^{+3.13}_{-2.92}$ & $0.50^{+0.09}_{-0.10}$ & $41.03^{+6.31}_{-6.13}$ & $3.14^{+0.11}_{-0.10}$ & $8.72^{+1.18}_{-1.03}$ & 1.91 & $1470.94^{+6.83}_{-6.85}$ & $22.49^{+0.24}_{-0.24}$ \\
52446 & $0.98^{+0.01}_{-0.01}$ & $85.89^{+2.99}_{-2.78}$ & $0.50^{+0.13}_{-0.15}$ & $36.90^{+8.16}_{-7.76}$ & $2.50^{+0.04}_{-0.04}$ & $14.23^{+0.93}_{-0.86}$ & 0.67 & $1469.96^{+7.42}_{-7.41}$ & $56.82^{+0.33}_{-0.33}$ \\
52446 & $0.96^{+0.01}_{-0.01}$ & $93.81^{+3.35}_{-3.08}$ & $0.52^{+0.09}_{-0.10}$ & $46.57^{+7.47}_{-7.19}$ & $3.25^{+0.12}_{-0.12}$ & $10.88^{+1.63}_{-1.40}$ & 1.36 & $1524.52^{+7.21}_{-7.23}$ & $27.89^{+0.31}_{-0.31}$ \\
52450 & $0.92^{+0.01}_{-0.01}$ & $90.92^{+4.07}_{-3.71}$ & $0.45^{+0.14}_{-0.16}$ & $23.07^{+6.04}_{-5.64}$ & $2.34^{+0.06}_{-0.06}$ & $9.93^{+0.98}_{-0.88}$ & 0.90 & $1217.35^{+6.49}_{-6.49}$ & $40.25^{+0.37}_{-0.37}$ \\
52452 & $0.89^{+0.01}_{-0.01}$ & $95.48^{+4.68}_{-4.27}$ & $0.48^{+0.11}_{-0.13}$ & $23.39^{+5.43}_{-5.16}$ & $2.36^{+0.06}_{-0.06}$ & $9.66^{+1.00}_{-0.90}$ & 0.85 & $1144.23^{+6.26}_{-6.24}$ & $36.27^{+0.34}_{-0.34}$ \\
52453 & $0.91^{+0.01}_{-0.01}$ & $82.91^{+3.65}_{-3.31}$ & $0.51^{+0.12}_{-0.14}$ & $23.02^{+5.40}_{-5.06}$ & $2.20^{+0.05}_{-0.05}$ & $7.71^{+0.70}_{-0.63}$ & 1.06 & $1085.81^{+5.73}_{-5.74}$ & $36.16^{+0.31}_{-0.31}$ \\
52453 & $0.90^{+0.01}_{-0.01}$ & $75.76^{+4.94}_{-4.21}$ & $0.55^{+0.19}_{-0.24}$ & $17.46^{+6.65}_{-5.83}$ & $2.33^{+0.03}_{-0.03}$ & $17.42^{+1.15}_{-1.06}$ & 0.41 & $931.70^{+5.59}_{-5.58}$ & $67.89^{+0.34}_{-0.34}$ \\
\hline
\end{tabular}

\vspace{0.2cm}
\begin{flushleft}
MJD: Modified Julian Date. 
$T_{\rm in}$: inner disk temperature from \texttt{diskbb}. 
$N_{\rm disk}$: normalization of the \texttt{diskbb} component. 
$\sigma$: line width of the Gaussian component. 
$N_{\rm gauss}$: normalization of the Gaussian component. 
$\Gamma$: photon index of the \texttt{nthComp} component. 
$N_{\rm nth}$: normalization of the \texttt{nthComp} component. 
$\chi^2$: reduced chi-squared of the fit. 
$L_{\rm Disk}$: unabsorbed disk luminosity in $3$–$25$ keV. 
$L_{\rm Compton}$: unabsorbed Comptonized luminosity in $3$–$25$ keV. Other details can be found in Sec. \ref{sec:fitting}.

Table~\ref{tab:1543spectral} is published in its entirety in the electronic version of the journal. A portion is shown here for guidance.
\end{flushleft}
\end{table*}

\begin{figure}
\vspace{-0.5cm}
    \centering  
    \includegraphics[width=0.5\columnwidth]{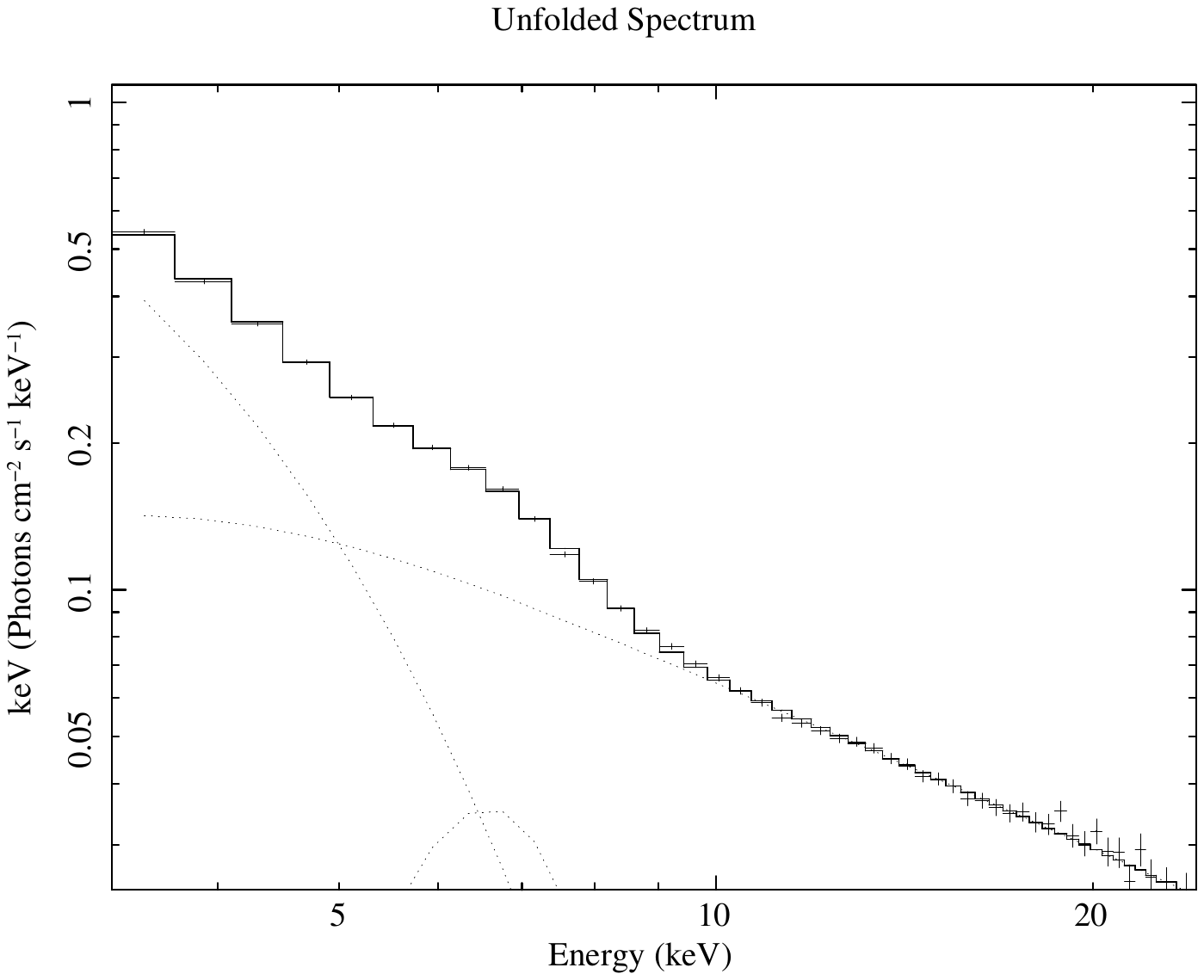}   
    \caption{An example of spectral fitting for GX 339-4 on MJD 53232.}  
    \label{fig:339fitting}  
\end{figure}


\bsp	
\label{lastpage}
\end{document}